\documentclass[twocolumn,twocolappendix]{aastex631}

\usepackage[utf8]{inputenc}
\usepackage{amsmath, amsfonts, amssymb, graphics,graphicx, wrapfig, nicefrac}
\usepackage{epsfig}
\usepackage{epstopdf}
\usepackage{multirow}
\usepackage{graphicx}
\usepackage{rotating}
\usepackage{amsthm}
\usepackage{mathrsfs}

\shorttitle{Chemistry in IC~348}
\shortauthors{Luo et al.}
\graphicspath{{./}{figures/}}

\begin{document}




\title{Dependence of Chemical Abundance on the Cosmic Ray Ionization Rate in IC~348}

\correspondingauthor{Zhi-Yu Zhang, Gan Luo}
\email{zzhang@nju.edu.cn, luogan@nju.edu.cn}

\author[0000-0002-1583-8514]{Gan Luo}
\affiliation{School of Astronomy and Space Science, Nanjing University, Nanjing 210093, People’s Republic of China}
\affiliation{Key Laboratory of Modern Astronomy and Astrophysics (Nanjing University), Ministry of Education, Nanjing 210093, People’s Republic of China}

\author[0000-0002-7299-2876]{Zhi-Yu Zhang}
\affiliation{School of Astronomy and Space Science, Nanjing University, Nanjing 210093, People’s Republic of China}
\affiliation{Key Laboratory of Modern Astronomy and Astrophysics (Nanjing University), Ministry of Education, Nanjing 210093, People’s Republic of China}

\author[0000-0003-2733-4580]{Thomas G. Bisbas}
\affiliation{Research Center for Intelligent Computing Platforms, Zhejiang Laboratory, Hangzhou 311100, China}
\affiliation{I. Physikalisches Institut, Universit$\ddot{a}$t zu K$\ddot{o}$ln, Z$\ddot{u}$lpicher Stra$\beta$e 77, D-50923 K$\ddot{o}$ln, Germany}

\author[0000-0003-3010-7661]{Di Li}
\affiliation{CAS Key Laboratory of FAST, National Astronomical Observatories, Chinese Academy of Sciences, Beijing 100101, China}
\affiliation{University of Chinese Academy of Sciences, Beijing 100049, China}
\affiliation{NAOC-UKZN Computational Astrophysics Centre, University of KwaZulu-Natal, Durban 4000, South Africa}

\author[0000-0002-2169-0472]{Ningyu Tang}
\affiliation{Department of Physics, Anhui Normal University, Wuhu, Anhui 241002, China}

\author[0000-0001-6106-1171]{Junzhi Wang}
\affiliation{School of Physical Science and Technology, Guangxi University, Nanning 530004, People's Republic of China}

\author[0000-0002-5683-822X]{Ping Zhou}
\affiliation{School of Astronomy and Space Science, Nanjing University, Nanjing 210093, People’s Republic of China}
\affiliation{Key Laboratory of Modern Astronomy and Astrophysics (Nanjing University), Ministry of Education, Nanjing 210093, People’s Republic of China}

\author[0000-0003-3948-9192]{Pei Zuo}
\affiliation{Kavli Institute for Astronomy and Astrophysics, Peking University, Beijing, 5 Yiheyuan Road, Haidian District, Beijing 100871, China}
\affiliation{International Centre for Radio Astronomy Research (ICRAR), University of Western Australia, Crawley, WA 6009, Australia}

\author[0000-0003-0355-6875]{Nannan Yue}
\affiliation{Kavli Institute for Astronomy and Astrophysics, Peking University, Beijing, 5 Yiheyuan Road, Haidian District, Beijing 100871, China}

\author[0000-0002-0818-1745]{Jing Zhou}
\affiliation{School of Astronomy and Space Science, Nanjing University, Nanjing 210093, People’s Republic of China}
\affiliation{Key Laboratory of Modern Astronomy and Astrophysics (Nanjing University), Ministry of Education, Nanjing 210093, People’s Republic of China}

\author[0000-0002-2231-8381]{Lingrui Lin}
\affiliation{School of Astronomy and Space Science, Nanjing University, Nanjing 210093, People’s Republic of China}
\affiliation{Key Laboratory of Modern Astronomy and Astrophysics (Nanjing University), Ministry of Education, Nanjing 210093, People’s Republic of China}



\begin{abstract}

Ions (e.g., H$_3^+$, H$_2$O$^+$) have been used extensively to quantify the cosmic-ray ionization rate (CRIR) in diffuse sightlines. However, measurements of CRIR in low-to-intermediate density gas environments are rare, especially when background stars are absent. In this work, we combine molecular line observations of CO, OH, CH, and HCO$^+$ in the star-forming cloud IC~348, and chemical models to constrain the value of CRIR and study the response of the chemical abundances distribution. The cloud boundary is found to have an $A_{\rm V}$ of approximately 4\,mag. From the interior to the exterior of the cloud, the observed $^{13}$CO line intensities drop by an order of magnitude. 
The calculated average abundance of $^{12}$CO (assuming $^{12}$C/$^{13}$C = 65) is (1.2$\pm$0.9) $\times$10$^{-4}$, which increases by a factor of 6 from the interior to the outside regions. The average abundance of CH (3.3$\pm$0.7 $\times$ 10$^{-8}$) is in good agreement with previous findings in diffuse and translucent clouds ($A_{\rm V}$ $<$ 5\,mag). However, we did not find a decline in CH abundance in regions of high extinction ($A_{\rm V}\simeq$8\,mag) as previously reported in Taurus. By comparing the observed molecular abundances and chemical models, we find a decreasing trend of CRIR as $A_{\rm V}$ increases. The inferred CRIR of $\zeta_{cr}$ = (4.7$\pm$1.5) $\times$ 10$^{-16}$\,s$^{-1}$ at low $A_{\rm V}$ is consistent with H$^+_3$ measurements toward two nearby massive stars.

\end{abstract}



\section{Introduction} \label{sec:intro}

The formation and evolution of interstellar molecules, which regularize the gas properties, such as gas cooling and the ionization fraction, are critical for understanding the physical conditions (e.g., density, temperature) and the evolution of the star-formation process (e.g., ambipolar diffusion). The intensity of the ultraviolet (UV) radiation field and the cosmic-ray ionization rate (CRIR, $\zeta_\mathrm{cr}$\footnote{The parameter $\zeta_{\rm cr}$ represent the total CRIR of H$_2$, namely $\zeta_2$}) are two of the most important parameters that govern the formation and evolution of molecules \citep{Draine1978,Wolfire2010,Padovani2013,Grenier2015}. (Far-)UV (FUV) photons with energies of $h\nu >$ 6 eV emitted by young massive stars, can heat the dust, penetrate the envelope of the molecular cloud, and influence the chemistry of the interstellar gas through photoionization and/or photodissociation processes. In the cloud interior where massive stars are absent, the shielding from the interstellar UV field becomes sufficient enough to allow the formation of molecules such as H$_2$ and CO. Cosmic-rays (CRs) on the other hand, are found to penetrate the interstellar medium (ISM) at high column densities, where FUV photons have been severely attenuated. Thus, CRs play a major role in the ionization and heating of dense and starless regions \citep{Padovani2020}.


Unlike the UV radiation field that can be conveniently inferred from dust emission or line ratios \citep[e.g., C$^+$/CO,][]{Kaufman1999,Pineda2013}, measuring the CRIR value is not that straightforward and is strongly dependent on the methods and environments (see the review of \citet{Dalgarno2006} and references therein). The CRIR may cover a wide range of values (10$^{-18}$\,s$^{-1}$ to 10$^{-15}$\,s$^{-1}$ or above) in nearby molecular clouds. 
H$^+_3$ is the most commonly used CRIR tracer in diffuse molecular cloud regions, and also the most direct one due to its connection with CRs; it is produced by the CR ionization of H$_2$ and it is destroyed through reactions of abundant neutral species \citep[e.g., CO, O, N$_2$,][]{McCall1999,Geballe1999, Dalgarno2006,Indriolo2012}. The downside is that measurements using H$^+_3$ require the existence of background massive stars, which limit their usage on molecular clouds.

Other molecules that are relevant to the H$^+_3$ chemistry are considered as potential probes of CRIR when combined with chemical models, e.g., OH and HD \citep{Hartquist1978}, HCO$^+$ and DCO$^+$ \citep{Guelin1982, van2000, Caselli1998}, OH$^+$ and H$_2$O$^+$ \citep{Gerin2010b, Neufeld2017, Bialy2019}. However, most of the measurements from ions (e.g., H$^+_3$, OH$^+$, H$_2$O$^+$), which are along diffuse sightlines, are limited by the background massive stars. 
\citet{Caselli1998} proposed that the CRIR in the dense cloud can be inferred from the abundance ratios of DCO$^+$/HCO$^+$ and HCO$^+$/CO, although the deuterium fraction (DCO$^+$/HCO$^+$) may strongly depend on the initial conditions (e.g., ortho-to-para ratio, temperature) \citep{Shingledecker2016}. In addition, carbon-chain molecules (e.g., HC$_3$N, HC$_5$N, c-C$_3$H$_2$) have also been used to infer the CRIR with chemical models toward the envelope of the protocluster OMC-2 FIR~4 \citep{Fontani2017, Favre2018}. However, due to their relatively low abundance, deuterium species as well as the carbon-chain molecules can only be detected in high extinction regions, such as in dense cores. It is, thus, difficult to constrain the CRIR value in low-to-intermediate density gas when background massive stars are absent.

In this work, we introduce our analysis from both observations and chemical models in the low-to-intermediate density ($n_\mathrm{H}$ $\sim$ 10$^4$\,cm$^{-3}$) star-forming cloud, IC~348, which put constraints on the CRIR. IC~348 is a nearby \citep[$\sim$316\,pc,][]{Herbig1998} star-forming cloud as part of the giant molecular cloud (GMC) -- Perseus molecular cloud \citep[PMC,][]{Bally2008}. Compared to its neighborhood -- Taurus, PMC has a lower mass \citep[$\sim$10$^4$\,M$_\odot$, half of Taurus,][]{Bally2008}, but higher dust temperature \citep[10--36\,K,][]{Zari2016}, higher UV radiation field \citep[$\chi/\chi_0$ $\sim$ 40, where $\chi_0$ is in unit of Draine field,][]{Draine1978,Sun2008}, and higher star formation rate \citep[96 or 150\,M$_\odot$ Myr$^{-1}$,][]{Evans2009, Mercimek2017}. In particular, the eastern portion (e.g., IC~348) of PMC is located at the center of the Per OB2 association, which would be influenced by the complex effects of stellar wind, UV radiation field, and shocks due to the possibly past supernova explosions \citep{Bally2008,Bialy2021,Kounkel2022}. Recent $\gamma$-ray observations found an exceeded value than the expected upper limits toward PMC \citep{Albert2021}. A very low $^7$Li/$^6$Li isotope ratio ($\sim$2) has been reported toward a star -- HD 281159 in IC~348 than that in the solar system \citep[12.2,][]{Knauth2017}, in which the generation of $^6$Li can only be attributed to spallation reactions by CRs \citep{Reeves1970,Meneguzzi1971}. All these pieces of evidence indicate that the eastern portion of PMC, especially IC~348, may have a high CRIR value. Thus, it is a good site to investigate how the UV radiation field and CRs influence the chemical abundances.

To better characterize the behavior of molecular abundances with the physical conditions (e.g., density, CRIR), we have observed three molecules (OH, CH, and HCO$^+$) that are relevant to the CO chemistry. OH and CH are proposed as alternative tracers of molecular gas, especially in low-to-intermediate density gas \citep{Liszt2002, Cotten2012, Xu2016b, Li2018, Busch2021, Tang2021a}. OH 18\,cm thermal emission and absorption lines have been detected extensively toward dark clouds, diffuse/translucent clouds, and extended to the outskirts of the molecular clouds where CO emission is faint or undetectable \citep{Turner1979, Magnani1988, Wannier1993, Cotten2012, Li2018}. 
Given its high abundance and the preceding formation path in photodissociation regions (PDRs), OH has been considered one of the most appropriate tracers of low-to-intermediate density gas \citep{Liszt2002, Cotten2012, Xu2016a}. CH is one of the first molecules detected in the ISM with optical absorption lines toward massive stars \citep{Dunham1937,Swings1937,McKellar1940}. Due to the fairly constant abundance ($\sim$4$\times$10$^{-8}$) in low extinction regions ($A_{\rm V}$ $<$ 5\,mag) detected by both 9\,cm radio emission and optical absorption lines \citep{Liszt2002, Sheffer2008, Tang2021b}, CH has been frequently used to quantify the H$_2$ column density \citep{Gerin2010a, Jacob2019}. As one of the precursors of CO, CH could be equally important as OH to the formation of CO and heavy complex molecules in gas-phase chemistry \citep{van1988,Herbst1989}. With the observations of CO, OH, CH, and HCO$^+$ toward IC~348, we will attempt to reveal the formation route that governs the chemistry of CO (e.g., the fraction of the CO formation through OH or CH channels).

This paper is organized as follows. The observations and archival data used in this work are presented in Section \ref{sec:obs}. In Section \ref{sec:results}, we derive the column density and abundance of each molecule. The photodissociation region (PDR) modelling is presented in Section \ref{sec:model}. In Section \ref{sec:discussion}, we discuss the uncertainties in deriving the molecular abundance and constraining the CRIR in IC~348. The main results and conclusions are summarized in Section \ref{sec:conclusion}.

\section{Observations}\label{sec:obs}


\subsection{Arecibo OH and CH Observations}\label{sec:oh and ch data}

\begin{figure}
\includegraphics[width=1.0\linewidth]{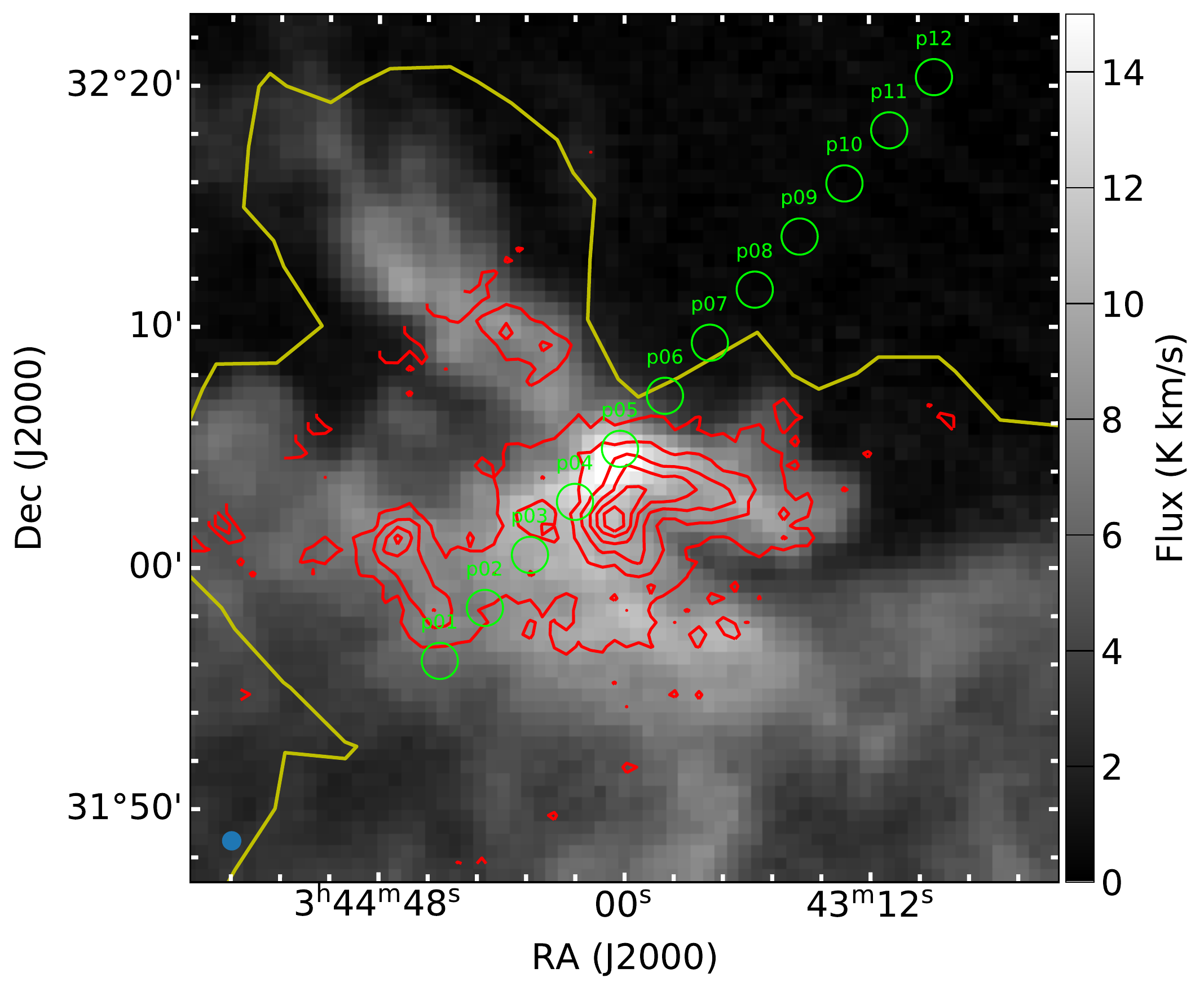}
\caption{The background gray map shows the $^{13}$CO (1--0) integrated intensity, in which the integrated velocity is in the range of 5 $\sim$ 12\,km\,s$^{-1}$. Red contours denote the HCO$^+$ integrated intensity at 0.2 K\,km\,s$^{-1}$ $\times$ (3, 5, 7, 9, 11, 13) with the same integrate range as that of $^{13}$CO (1--0). The 1.5$'$ green circles denote the 12 points of OH and CH observations. Yellow solid line outlines where the visual extinction ($A_{\rm V}$) is 4\,mag.} 
\label{fig:IC348}
\end{figure}

We have observed four OH $\Lambda$-doubling lines in the $^2\Pi_{3/2}$, $J$ = 3/2 energy level, using the L-band Wide receiver with the total power ON mode. The Project ID is A3224, and the observations were conducted from Jan. 2018 to Feb. 2018. The rest frequencies for the four transitions are 1612.2310, 1665.4018, 1667.3590, and 1720.5300\,MHz. The four OH transitions have been placed into four sub-bands, where each sub-band has a bandwidth of 3.125\,MHz. The spectral resolution of OH is 381\,Hz (equal to 0.069\,km\,s$^{-1}$ at 1.6\,GHz), and it is then smoothed to a velocity resolution of 0.14\,km\,s$^{-1}$ to obtain a higher signal-to-noise ratio. The integrated time is 1500\,s for OH spectra at each position, resulting in an rms noise level of $\sim$0.015\,K per 0.14\,km\,s$^{-1}$.
The three CH $\Lambda$-doubling lines in the $^2\Pi_{1/2}$, $J$ = 1/2 energy level were taken using the S-band high receiver with the total power ON mode. The rest frequencies for the three transitions are 3335.481, 3263.794, and 3349.193\,MHz, each band has a bandwidth of 3.125\,MHz. The spectra of CH are smoothed to a velocity resolution of 0.14\,km\,s$^{-1}$. The integrated time is 2100\,s for the CH spectra at each position, resulting in an rms noise level of $\sim$0.015\,K per 0.14\,km\,s$^{-1}$. 
The beam efficiency for OH and CH observations are 50$\%$ and 40$\%$, respectively. Both the OH and CH spectra were taken at 12 positions across the IC~348, as shown with green circles in Figure \ref{fig:IC348}.

\subsection{HCO$^+$\ Observations}\label{sec:hcop data}

The HCO$^+$\ $J$ = 1--0 (89.188\,GHz) observations were taken during May 2018 using the Delingha 13.7-m telescope with On-The-Fly (OTF) mode. The observed map is $\sim$40$'$ along the northwest-southeast direction and $\sim$10$'$ along the northeast-southwest direction, which covers the 12 Arecibo pointings. The integrated intensity distribution of HCO$^+$ is shown in red contours in Figure \ref{fig:IC348}. The spectral resolution is 61\,kHz, which corresponds to 0.21\,km\,s$^{-1}$ at 89\,GHz. The beam efficiency is 56$\%$ for HCO$^+$ observations. The datacube was re-gridded to a spatial resolution of 30$''$, and the rms noise level is $\sim$0.02\,K per 0.21\,km\,s$^{-1}$.

\subsection{Archive extinction data}\label{sec:extinction}
We use the 2MASS Final Perseus Extinction Map from the COMPLETE team \citep{complete2011} to estimate the total gas column density. The data values in this map are magnitudes of visual extinction ($A_{\rm V}$), which are made from 2MASS and NICER (Near Infrared Extinction-method Revisited). It was re-gridded to a spatial resolution of 2.5$'$. The $A_{\rm V}$ value ranges from $2-10\,{\rm mag}$ in the observed region.

\subsection{Archive H\,{\sc i} Data}\label{sec:hi data}
We used H\,{\sc i} spectra from the Galactic Arecibo L-Band Feed Array H\,{\sc i} (GLAFA-H\,{\sc i}) survey \citep{Peek2011}, in which the column density of H\,{\sc i} is calculated by integration over a velocity range of $-$400\,km\,s$^{-1}$ $<$ V $<$ 400\,km\,s$^{-1}$ under the assumption of $\tau \ll 1$. GALFA-H\,{\sc i} is a large-scale survey of Galactic interstellar medium in the 21\,cm line hyperfine transition of neutral hydrogen, undertaken with the Arecibo L-band Feed Array (ALFA) at the Arecibo Observatory. The survey covers a large area (13,000 $\mathrm{deg^2}$) with high spatial resolution ($\sim$4$'$), high spectral resolution (0.18\,km\,s$^{-1}$), and large bandwidth ($-$700\,km\,s$^{-1}$\ $<$ $\mathrm{V_{LSR}}$ $<$ +700\,km\,s$^{-1}$). The datacube was regridded to a spatial resolution of 1$'$), and the typical noise level is 80\,mK per 1\,km\,s$^{-1}$ channel. 

\subsection{FCRAO $^{12}$CO\ and $^{13}$CO\ ($J$ = 1-0) Data}\label{sec:co data}
The $^{12}$CO $J$ = 1--0 and $^{13}$CO $J$ = 1--0 observations were taken by the FCRAO 13.7-m telescope using the OTF mapping method during 2002-2004 \citep{Ridge2006}. The map center was at R. A. = $03^h 37^m 00^s$ and Dec. = $+31^\circ 49'00''$. The beam size is 46$"$, and the data were regridded to $\sim$23$''$, resulting in rms noise levels of 0.10 and 0.05\,K for $^{12}$CO and $^{13}$CO observations. The spectral resolutions are 39.7\,kHz for $^{12}$CO and 41.5\,kHz for $^{13}$CO, corresponding to a velocity resolution of 0.064 and 0.066\,km\,s$^{-1}$, respectively. The main beam efficiencies are 0.49 at 110\,GHz and 0.45 at 115\,GHz.

\subsection{JCMT $^{13}$CO\ ($J$ = 3-2) Data}\label{sec:co32 data}
The $^{13}$CO $J$ = 3--2 observations were taken by the HRAP imaging array in JCMT on Dec. 17, 2007. The image covers a 1100$''$ $\times$ 650$''$ region, which is centered at R. A. = $03^h 44^m 14^s$ and Dec. = $+31^\circ 49'49''$. The spectral resolution of the $^{13}$CO $J$ = 3-2 spectra is 61\,kHz (0.05\,km\,s$^{-1}$ at 330\,GHz), and it is rebinned to a spectral resolution of 0.15\,km\,s$^{-1}$. The beam size is 17.7$"$, and the beam efficiency is 0.66. The rms noise level is 0.7\,K per 0.05\,km\,s$^{-1}$ channel. Details of the observations can be found in \citet{Curtis2010}.

\section{Results}\label{sec:results}
\subsection{Line profiles}\label{sec:lines}

\begin{figure*}
\includegraphics[width=1.0\linewidth]{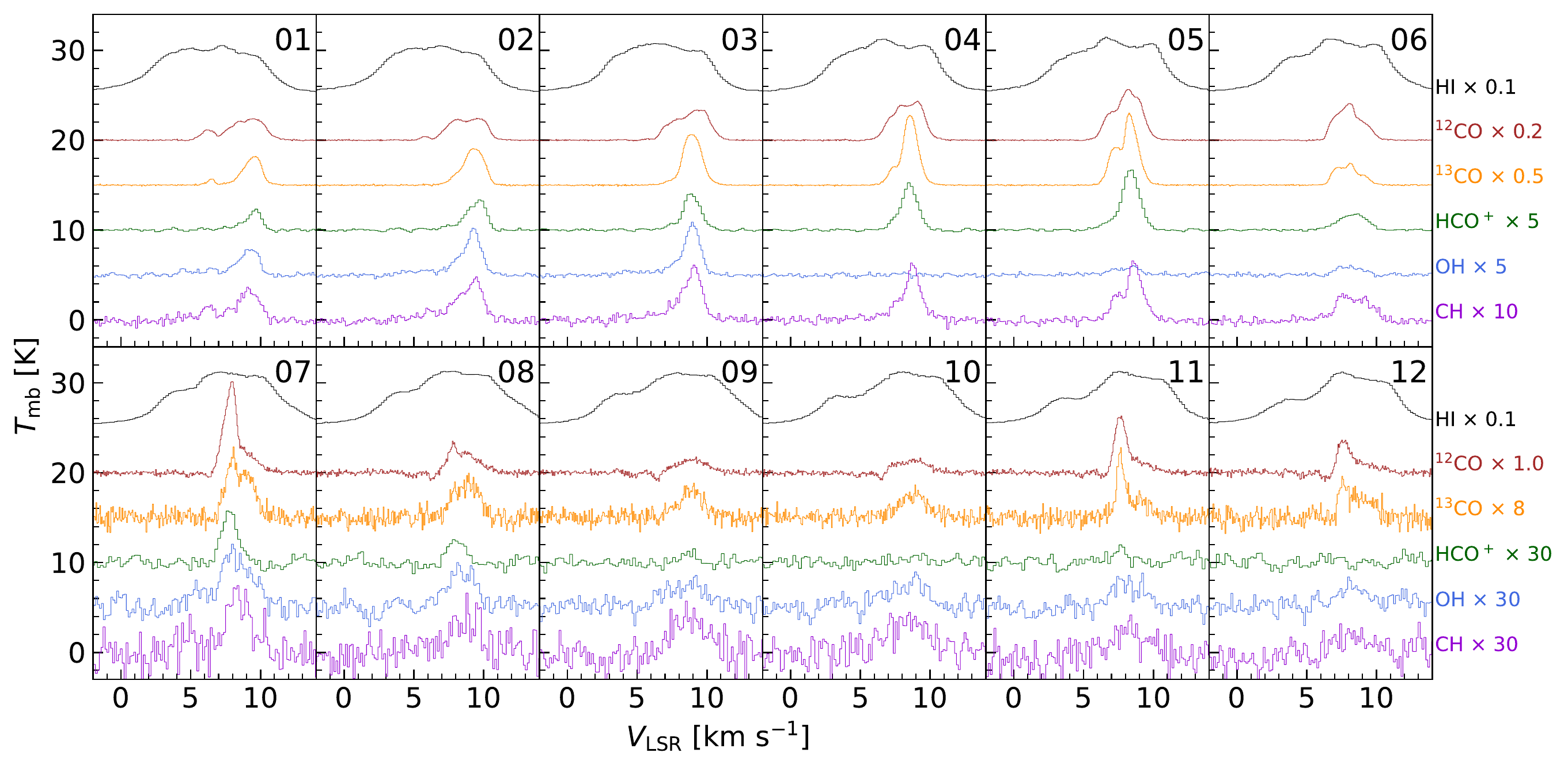}
\caption{The spectra of H\,{\sc i}, $^{12}$CO, $^{13}$CO, HCO$^+$, OH 1667 main line, and CH 3335 main line toward twelve points (numbers from 01 to 12 denote positions from p01 to p12). The scaling factors of H\,{\sc i}, $^{12}$CO, $^{13}$CO, HCO$^+$, OH, and CH are listed in the right panel of this figure. To avoid overlapping, the spectra of H\,{\sc i}, $^{12}$CO, $^{13}$CO, HCO$^+$, and OH have been shifted upwards with an interval of 5 in y-axis. The x- and y-axis denote the velocity range in km\,s$^{-1}$ and the brightness temperature, respectively.}
\label{fig:lines}
\end{figure*}

The twelve points of the OH and CH observations are shown as green circles in Figure \ref{fig:IC348}, which are labeled as p01 to p12. Figure \ref{fig:lines} shows the spectra of H\,{\sc i}, $^{12}$CO (1--0), $^{13}$CO (1--0), HCO$^+$ (1--0), OH, and CH toward the twelve pointings, in which $^{12}$CO (1--0), $^{13}$CO (1--0), and HCO$^+$ (1--0) data are spatially re-binned to 3$'$ at each pointing\footnote{The spatial resolution of H\,{\sc i} (4$'$) is larger than OH, we assume that the spatial fluctuation of HI is flat in 4$'$ beam.} (the representative re-binned spatial distribution of $^{13}$CO (1--0), HCO$^+$ (1--0), and extinction map are presented in Appendix \ref{sec:figures}). If we define the cloud boundary the position where the intensity of $^{12}$CO (1--0) or $^{13}$CO (1--0) drops significantly (e.g., by an order of magnitude for $^{13}$CO, see the spectra in Figure~\ref{fig:lines}), the boundary roughly coincide with the $A_{\rm V}$ of 4\,mag (see Figure~\ref{fig:IC348}). Then, p01 to p06 are located in the interior of cloud regions (ICRs) and p07 tp p12 are located in the exterior of cloud regions (ECRs).

The central line velocity ranges from 6.5--9.8\,km\,s$^{-1}$ for $^{13}$CO and HCO$^+$, 6.1--9.5\,km\,s$^{-1}$ for OH 1667\,MHz, and 4.6--9.4\,km\,s$^{-1}$ for CH 3335\,MHz. Self-absorption signatures of H\,{\sc i} and $^{12}$CO spectra exist in the velocity range of $\sim$8--10\,km\,s$^{-1}$ toward p01 to p06, which corresponds to the emission peaks of $^{13}$CO and HCO$^+$. Spectra of $^{13}$CO, HCO$^+$, OH, and CH show velocity gradients in the direction from p01 to p06, with red-shifted velocities ($\sim$10\,km\,s$^{-1}$) at p01 and blue-shifted velocities ($\sim$8\,km\,s$^{-1}$) at p06. 

\subsection{Column density calculation with LTE}\label{sec:column density}

We now derive the column densities of $^{13}$CO, HCO$^+$, OH, and CH under the assumption of local thermal equilibrium (LTE).
The column density can be written as a function of optical depth \citep[$\mathrm{\tau_\nu}$,][]{Mangum2015}:
\begin{equation}
N_{tot} = \frac{3h}{8{\pi}^3\left | \mu_{lu} \right |^2} \frac{Q_{rot}}{g_{u}} \frac{e^{\frac{E_u}{kT_{ex}}}} {e^{\frac{h\nu}{kT_{ex}}}-1}  \int \tau_\nu d\upsilon,
\label{eq:n_tot}
\end{equation}
where $\mathrm{\left | \mu_{lu} \right |^2}$ is the dipole matrix element, textbf{u and l represent the upper and lower energy level, respectively}. $\mathrm{Q_{rot}}$ is the rotational partition function, $\mathrm{g_u}$ is the degeneracy of upper energy level, $\mathrm{E_u}$ is the energy of upper energy level, and $T_\mathrm{{ex}}$ is the excitation temperature.
According to the radiative transfer equation, the brightness temperature $T_\mathrm{b}$ is given by the expression, 
\begin{equation}
T_{b} = \eta [J(T_{ex})-J(T_{bg})](1-e^{-\tau_\nu}),
\label{eq:t_mb}
\end{equation}
where $\eta$ is the beam filling factor (assumed to be 1 in the following calculation). $T_\mathrm{bg}$ is the background continuum temperature, which should be equal to the cosmic microwave background (CMB, 2.73\,K at z = 0) for $^{13}$CO and HCO$^+$ observations. For the low frequency OH and CH observations, the Galactic synchrotron background should be taken into account. Note that the temperature terms in equation \ref{eq:t_mb} should be in the form of the Planck function: $J\mathrm{(T)}$ = $(h\nu/k)/(\exp(h\nu/kT)-1)$ at the millimeter wavelengths.

With the optically thin assumption ($\mathrm{\tau_\nu \ll 1}$), equation \ref{eq:n_tot} can be rewritten in the form of $T_\mathrm{b}$:
\begin{equation}
N_{thin} = \frac{3h}{8{\pi}^3\left | \mu_{lu} \right |^2} \frac{Q_{rot}}{g_{u}} \frac{e^{\frac{E_u}{kT_{ex}}}} {e^{\frac{h\nu}{kT_{ex}}}-1} \frac{\int T_{b} d\upsilon}{J(T_{ex})-J(T_{bg})} .
\label{eq:n_thin}
\end{equation}

For each transition, the $\mathrm{\left | \mu_{lu} \right |^2}$, $\mathrm{E_u}$, and rest frequency $\nu$ are taken from the CDMS database\footnote{https://cdms.astro.uni-koeln.de/cdms/portal/} \citep{Muller2001,Muller2005}.

\subsubsection{$^{13}$CO}\label{sec:co}

The self-absorption components of $^{12}$CO are well-matched with the $^{13}$CO components in velocity, meaning that the $^{12}$CO transitions are optically thick (especially in the ICRs). Thus, the $^{12}$CO emission is less reliable in deriving the column density. 
We use two different methods to derive the excitation temperature of $^{13}$CO: the first one is by using $^{12}$CO (1--0) (hereafter, $T^1_\mathrm{{ex}}$), and the second one by using $^{13}$CO (1--0) and (3--2) (hereafter, $T^2_\mathrm{{ex}}$). More details of the excitation temperature can be found in Appendix \ref{sec:tex}.

Once we obtain the excitation temperature, the column density can be derived from the emission line of the $^{13}$CO (1--0) transition ($T_\mathrm{13}$) by adding an optical correction factor:
\begin{equation}
N_{^{13}CO} = \frac{\tau_{13}}{1-e^{-\tau_{13}}} \times N_{thin,^{13}CO},
\label{eq:n_13co}
\end{equation}
where the optical depth of $^{13}$CO ($\tau_{13}$) is in the form of:
\begin{equation}
\tau_{13} = -ln (1 - \frac{T_{13}}{J(T_{ex})-J(T_{bg})}).
\label{eq:tau13}
\end{equation}

If we substitute all the constants in equation \ref{eq:n_thin}, a convenient form of equation \ref{eq:n_13co} is:
\begin{equation}
\begin{matrix}
N_{^{13}CO} = 6.57 \times 10^{14} \frac{\tau_{13}}{1-e^{-\tau_{13}}} Q_{rot}e^{\frac{5.288}{T_{ex}}} \left [ e^{\frac{5.288}{T_{ex}}}-1 \right ]^{-1} \\
\frac{\int T_{13} d\upsilon ({\rm km\,s^{-1}}) }{J(T_{ex})-0.89} \ [{\rm cm^{-2}}].
\end{matrix}
\end{equation}

The resultant column density of $^{13}$CO ranges from (5.2$\pm$0.9)$\times$10$^{14}$ to (8.95$\pm$0.02)$\times$10$^{16}$\,cm$^{-2}$ (if adopting $T^1_\mathrm{{ex}}$) or (4.94$\pm$0.02)$\times$10$^{16}$\,cm$^{-2}$ (if adopting $T^2_\mathrm{{ex}}$), which spans two orders of magnitude across this region. However, the calculation under LTE does not always make sense in our situation (see Appendix \ref{sec:tex}). In the following analysis of this paper, we adopt the calculation of column density from non-LTE code RADEX \citep{Van2007}, which are presented in Appendix \ref{sec:radex}. The column density of $^{13}$CO from RADEX is in the range of (0.38$\pm$0.03 $\sim$ 38.9$\pm$0.1) $\times$ 10$^{15}$\,cm$^{-2}$. 


\subsubsection{HCO$^+$}\label{sec:hcop}


If we substitute all the constants in equation \ref{eq:n_thin}, the column density of HCO$^+$ as a function of $T_\mathrm{{ex}}$ in the optical thin assumption can be written as:
\begin{equation}
\begin{matrix}
N_{\rm HCO^+} = 5.30 \times 10^{11} \ Q_{rot}e^{\frac{4.28}{T_{ex}}} \left[ e^{\frac{4.28}{T_{ex}}}-1 \right ]^{-1} \\ \frac{\int T_{HCO^+} d\upsilon ({\rm km\,s^{-1}}) }{J(T_{ex})-1.13} \ [{\rm cm^{-2}}].
\end{matrix}
\label{eq:hcop}
\end{equation}

With only one transition of HCO$^+$, one can hardly obtain a reliable excitation temperature. 
Here, we roughly adopt the values of $T_\mathrm{{ex}}$ of HCO$^+$ as 4\,K in the following calculations. The derived column density of HCO$^+$ ranges from (1.5$\pm$0.3)$\times$10$^{11}$ to (4.09$\pm$0.02)$\times$10$^{12}$\,cm$^{-2}$.



\subsubsection{OH}\label{sec:oh}

The excitation condition of OH is strongly dependent on the environment. Non-thermal excitation (e.g., shocks) or FIR pumping mechanisms may lead to an anomalous hyperfine line intensity ratio \citep{Sivagnanam2004,Xu2016a}. In our observations, 6 out of 12 pointings have both OH 1665 and 1667 detections, in which the OH 1665/1667 line ratio (0.5$\sim$0.8) is close to the line ratio in LTE ($\sim$0.6). Thus, we can write the column density of OH as a function of the optical depth of the 1667\,MHz satellite line in LTE assumption \citep{Knapp1973, Dickey1981, Liszt1996}:
\begin{equation}
N_{OH} = 2.24 \times 10^{14} \ T_{ex} \int \tau_{1667} d\upsilon ({\rm km\,s^{-1}}) \ [{\rm cm^{-2}}].
\end{equation}

Assuming that the OH 1667\,MHz line is optically thin, the above equation can be written as:
\begin{equation}
N_{OH} = 2.24 \times 10^{14} \frac{T_{ex}}{T_{ex}-T_{bg}} \int T_{1667} d\upsilon ({\rm km\,s^{-1}}) \ [{\rm cm^{-2}}].
\end{equation}

Here, the Galactic background emission is estimated from the 408\,MHz all-sky continuum survey through the equation: $T_\mathrm{bg} = T_\mathrm{CMB} + T_{408}(\nu/408)^{-2.8}$ \citep{Haslam1982}. The continuum emission in our target region is 27\,K at 408\,MHz, resulting in a $T_\mathrm{bg}$ of 3.3\,K for OH and 2.8\,K for CH. In our calculation, we adopt the value of $T_\mathrm{ex} - T_\mathrm{CMB}$ = 2\,K as in the OH survey by \citet{Li2018} and \citet[][]{Tang2021a}. 
Thus, the resultant column density of OH ranges from (1.1$\pm$0.2) $\times 10^{14}$\,cm$^{-2}$ to (1.1$\pm$0.1) $\times 10^{15}$\,cm$^{-2}$.

\subsubsection{CH}\label{sec:ch}

The emission line of CH $J$ = 1/2 $\Lambda$-doublet lines in various astrophysical environments shows ubiquitous level inversion, even toward strong continuum sources \citep[e.g., H\,{\sc ii} regions, quasars,][]{Rydbeck1976, Liszt2002, Gerin2016, Jacob2021, Tang2021b}. The CH lines have a negative excitation temperature and are always optically thin ($\left | \tau \right | < 0.01$). The critical density is a few cm$^{-3}$ for $J$ = 1/2 $\Lambda$-doublet transitions. In contrast, the excited state ($J$ = 3/2) has a critical density as high as 10$^6$\,cm$^{-3}$. Considering that the typical gas density of molecular cloud should be around $\sim$10$^4$\,cm$^{-3}$, most of the CH molecules are at the $J$ = 1/2 level.

The column density of CH can be obtained through the 3335\,MHz main line \citep{Rydbeck1976, Magnani1989}:
\begin{equation}
N_{CH} = 2.82 \times 10^{14} \frac{T_{ex}}{T_{ex}-T_{bg}} \int T_{3335} d\upsilon ({\rm km\,s^{-1}}) \ [{\rm cm^{-2}}].
\end{equation}
The excitation temperature of the 3335\,MHz transition in previous works was frequently adopted as -15\,K \citep{Rydbeck1976} or -60\,K \citep{Genzel1979}. In this work, we adopted a typical $T_\mathrm{{ex}}$ value of -15\,K. Though the value of $T_\mathrm{{ex}}$ could vary by a factor of 4, however, since $\left | T_\mathrm{ex} \right | \geq T_\mathrm{bg}$, the variation of column density from $T_\mathrm{{ex}}$ is within 14\%. The resultant column density of CH ranges from (7$\pm$2) $\times 10^{13}$\,cm$^{-2}$ to (2.8$\pm$0.2) $\times 10^{14}$\,cm$^{-2}$.



\subsection{Abundance}\label{sec:abundance}

\begin{figure*}
\includegraphics[width=1.0\linewidth]{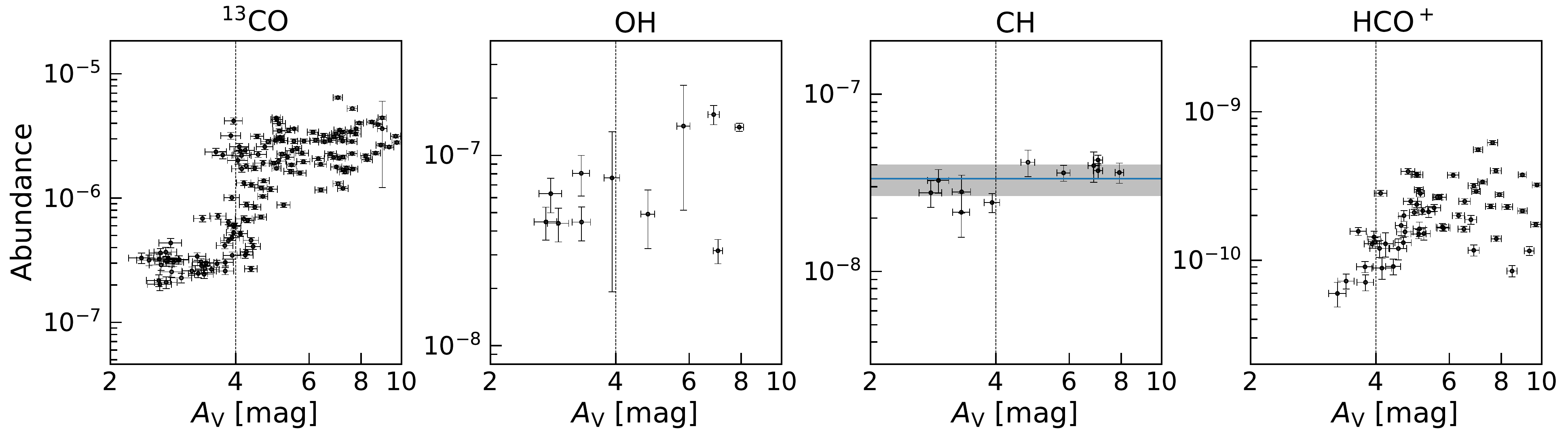}
\caption{From left-to-right: the abundances of $^{13}$CO, OH, CH, and HCO$^+$ versus $A_{\rm V}$. The black dashed lines denote $A_{\rm V}$ = 4\,mag. In the third panel, the blue horizontal line is the average abundance of CH, and the gray region outlines the 1\,$\sigma$ dispersion of CH abundance.}
\label{fig:abundance}
\end{figure*}

In this section, we estimate the relative abundance (column density ratio) of $^{13}$CO, HCO$^+$, OH, and CH with respect to H$_2$ (Figure \ref{fig:abundance}). The H$_2$ column density is derived by adopting  $N_{\rm H_2}=(N_\mathrm{H}-N_\mathrm{H\,{\sc I}})/2$, where $N_\mathrm{H}$ is the total H-nucleus column density and $N_\mathrm{H\,{\sc I}}$ is the H\,{\sc i} column density. The total H column density is converted from the visual extinction ($A_{\rm V}$) with the canonical conversion factor $N_\mathrm{H}/A_{\rm V}=1.8 \times 10^{21}$\,atoms\,cm$^{-2}$\,mag$^{-1}$ \citep{Bohlin1978, Predehl1995}. 

\subsubsection{Abundance of $^{13}$CO}\label{sec:abundance of co}


The abundance of $^{13}$CO ranges from (2.0$\pm$0.2) $\times 10^{-7}$ to (6.4$\pm$0.2) $\times 10^{-6}$ in IC~348, which increases by over an order of magnitude as the increasing $A_{\rm V}$. A clear dichotomy is found between ECRs (which is roughly outlined by $A_{\rm V}$ $<$ 4\,mag) and ICRs ($A_{\rm V}$ $>$ 4\,mag), in which the average abundance of the latter (2.3$\pm$1.1 $\times$10$^{-6}$) is more than six times higher than that of the former (3.8$\pm$3.8 $\times$10$^{-7}$). By assuming a constant $^{12}$C/$^{13}$C isotopic ratio of 65 for nearby clouds \citep{Langer1990, Langer1993, Milam2005, Giannetti2014}, the resultant abundance of $^{12}$CO ranges from (1.3$\pm$0.1) $\times 10^{-5}$ to (4.2$\pm$0.1) $\times 10^{-4}$ (in the following, we use CO instead of $^{12}$CO if there are no special circumstances). The mean value of CO abundance is (1.2$\pm$0.9) $\times$10$^{-4}$, which is similar to the canonical value found in nearby molecular clouds \citep[$\sim 1 \times 10^{-4}$,][]{Frerking1982,Blake1987,Pineda2010}.

\subsubsection{Abundance of HCO$^+$}\label{sec:abundance of hcop}

The abundance of HCO$^+$ is higher in the ICRs than that of ECRs, which ranges from (6$\pm$1) $\times$10$^{-11}$ to (6.2$\pm$0.2) $\times$10$^{-10}$. The HCO$^+$ abundance increases as the increasing $A_{\rm V}$. 
The mean value of the abundance of HCO$^+$ is (2$\pm$1)$\times$10$^{-10}$ across this region. This value is similar to that of measurements from protostellar envelopes \citep[in the order of 10$^{-10}$,][]{Jorgensen2004, Aikawa2021}, PGCCs \citep[10$^{-12}$ $\sim$ 10$^{-9}$,][]{Yuan2016, Wakelam2021, Yi2021}, and recent individual pointing observations toward Perseus molecular cloud \citep[$10^{-10} \sim 3\times10^{-9}$,][]{Baras2021}. The abundance of HCO$^+$ have large scatter (by a factor of 8) in the high extinction regions ($A_{\rm V}$ $>$6\,mag). We note that similar results have been reported by \citet{Baras2021}, in which the abundance of HCO$^+$ could change dramatically (by two orders of magnitude) toward different line-of-sights (LOSs), even though they share the same $A_{\rm V}$ value.  

\subsubsection{Abundance of OH}\label{sec:abundance of oh}

The abundance of OH is in the range of (3.2$\pm$0.5) $\times 10^{-8}$ to (1.6$\pm$0.2) $\times 10^{-7}$. The abundance of OH does not change much below $A_{\rm V}$ of 4\,mag, while it slightly increases by a factor of 3 at a higher $A_{\rm V}$. The average abundance of OH is (8$\pm$5) $\times$ 10$^{-8}$, which is in the range of recent OH surveys of 141 molecular clouds by \citet[5.7$\times$10$^{-8}$ $\sim$ 4.8$\times$10$^{-6}$,][]{Tang2021a}, and similar to that in diffuse and translucent clouds \citep[10$^{-7}$,][]{Liszt1996, Weselak2010, Nguyen2018, Li2018}. 

\subsubsection{Abundance of CH}\label{sec:abundance of ch}

The abundance of CH ($f_\mathrm{CH}$ = (3.3$\pm$0.7) $\times$ 10$^{-8}$) is nearly constant in our observations. This value is in good agreement with the value found in diffuse and translucent clouds ($A_{\rm V}$ $<$ 5\,mag) by radio emission observations toward extragalactic continuum sources \citep[(4.3$\pm$1.9) $\times$ 10$^{-8}$,][]{Liszt2002}, optical absorption observations toward stars \citep[$3.5^{+2.1}_{-1.4} \times 10^{-8}$,][]{Sheffer2008}, and recent radio ON--OFF observations against quasars \citep[(3.5$\pm$0.4) $\times$ 10$^{-8}$,][]{Tang2021b}. 

We note that the correlation between $N_\mathrm{CH}$ and $N_{\rm H_2}$ keeps linear at log ($N_{\rm H_2}/{\rm cm^{-2}}$) $<$ 21 \citep[spiral arm clouds,][]{Qin2010} or log ($N_{\rm H_2}/{\rm cm^{-2}}$) $<$ 21.5 \citep[$A_{\rm V}$ $<$ 5\,mag,][]{Mattila1986, Sakai2012}, while there is a noteworthy decline on the CH abundance above the $N_{\rm H_2}$ value (by an factor of 2$\sim$10). In addition to the differences in deriving the $N_{\rm H_2}$ by \citet{Qin2010}, however, the constant abundance of CH maintains up to log ($N_{\rm H_2}/{\rm cm^{-2}}$) $\sim$ 21.9 ($A_{\rm V}$ = 8\,mag) in our work.

\subsection{UV radiation field}\label{sec:uv}

As can be seen in Figure \ref{fig:abundance}, the $A_{\rm V}$ value ($\sim$4\,mag) for which the abundance of CO reaches a higher average value, is higher than the $A_{\rm V}$ ($\sim$2\,mag) that is found under usual ISM conditions \citep[$\chi/\chi_0$ = 10$\sim$30, $n_{\rm H}$ $\sim$ 10$^3$\,cm$^{-3}$,][]{Wolfire2010,Bolatto2013}. At low $A_{\rm V}$ regions, the main destruction of CO is photodissociation by UV photons. The low CO abundance at $A_{\rm V}$ $<$ 4\,mag may indicate unusual UV radiation field and chemistry in IC~348 (see Section \ref{sec:model} for more discussion).

Here, we present a rough estimation of the FUV intensity ($G_0$). If we ignore the heating of cosmic rays, and the heating outside the FUV band is equal to the FUV, the observed total infrared continuum intensity ($F_\mathrm{TIR}$) can be written as \citep{Kaufman1999,Kramer2005}:
\begin{equation}
F_{\rm TIR} = 2 \times 1.3 \times 10^{-4} G_0 \ [{\rm ergs \ cm^{-2} \ s^{-1} \ sr^{-1}}].
\end{equation}

The $F_\mathrm{TIR}$ at each position is estimated by integrating the spectral energy distribution (SED) fitting from 1\,$\mu$m to 1000\,$\mu$m. The emission at each observed frequency can be modelled by two modified blackbody functions \citep{Planck2014, Zari2016}:
\begin{equation}
I_{\nu} = \tau_{\nu_1}(\frac{\nu}{\nu_1})^{\beta_1}B_{\nu}(T_1) + \tau_{\nu_2}(\frac{\nu}{\nu_2})^{\beta_2}B_{\nu}(T_2),
\end{equation}
where the first and second items represent the cold ($T_1$) and warm ($T_2$) dust components, respectively. $\tau_{\nu}$ is the optical depth at reference frequency $\nu$, and $B_{\nu}(T)$ is the Planck function. The reference frequencies are set as $\nu_1$ = 1.2\,THz and $\nu_2$ = 10\,THz. To simplify the calculation, we set the spectral index $\beta_1$ = $\beta_2$ = 2.

\begin{figure}
\includegraphics[width=1.0\linewidth]{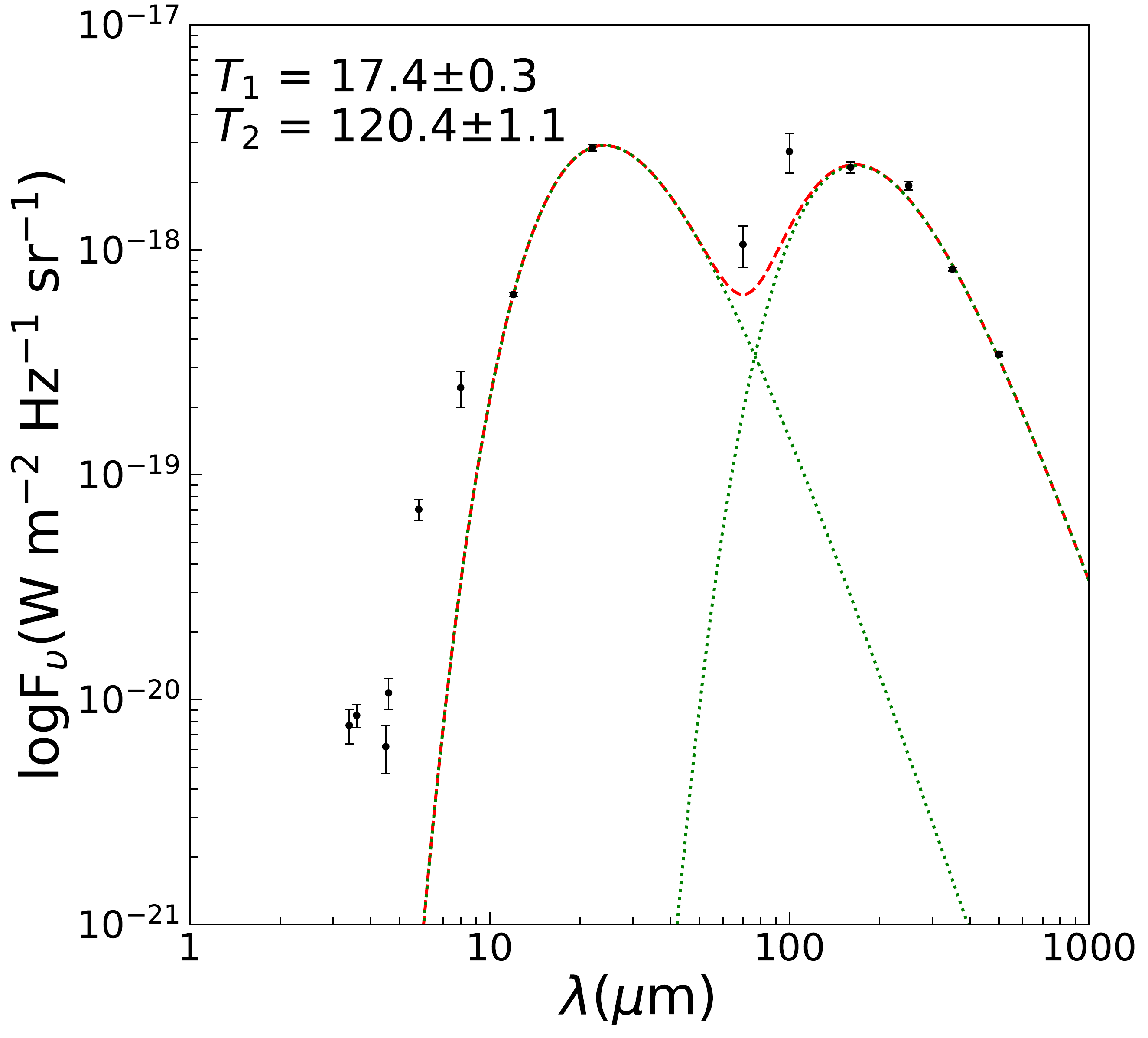}
\caption{The SED fitting (red dashed line) with two modified blackbodies toward p05. Each of the dust component is denoted with green dotted line. $T_1$ and $T_2$ represent cold dust temperature and warm dust temperature, respectively.}
\label{fig:sed}
\end{figure}

The data used for SED fitting are taken from $Spitzer$ \citep[3.6, 4.5, 5.8, and 8\,$\mu$m,][]{Werner2004,Fazio2004}, $WISE$ \citep[3.4, 4.6, 12, and 22\,$\mu$m,][]{Wright2010}, and $Herschel$ \citep[70, 100, 160, 250, 350, 500\,$\mu$m,][]{Pilbratt2010,Poglitsch2010,Griffin2010}. The $Herschel$ data were part of the Gould Belt Survey \citep{Andre2010}.
A representative SED fitting result toward p05 is shown in Figure \ref{fig:sed}.

We note that the observed flux at 70 and 100\,$\mu$m is higher than the SED fitting, which may be caused by the confusing emission from very small grains (VSGs, \citep[VSGs, grain radius ``r" $<$ 0.1\,$\mu$m,][]{Dwek1997, Li2001}. The emission from VSGs at 70\,$\mu$m in Perseus account for 70\% of the total flux at 17\,K and up to 90\% at 14\,K \citep{Schnee2005, Schnee2008}. The emission of VSGs is not expected to contribute much to wavelength $\lambda > 100 \mu$m \citep{Li2001}.

The resultant FUV radiation field in the whole 0.6$^\circ$ $\times$ 0.6$^\circ$ region is $\chi/\chi_0$ = 100$\pm$14 ($\chi_0$ = 1.7 $G_0$). This value is $\sim$2 times higher than the results by \citet{Sun2008}. We note that in \citet{Sun2008}, the $IRAS$ 60 and 100\,$\mu$m data were only used, which may underestimate the TIR intensity.

\section{Photodissociation Region Modelling}\label{sec:model}

In Section \ref{sec:uv} it was shown that the cloud environmental conditions are different from that of nearby GMCs. To better understand the chemical evolution of the observed species, the observational results are compared with chemical models under specific ISM environmental parameters (e.g., density, FUV intensity, CRIR). The chemical models are performed using the publicly available code {\sc 3d-pdr}\footnote{https://uclchem.github.io/3dpdr/} \citep{Bisbas2012}.
{\sc 3d-pdr} is an astrochemical code treating photodissociation regions of arbitrary density distributions, both in one- and three-dimensions. It performs iterations over thermal balance and converges once the total heating is equal to the total cooling to within a user-defined tolerance parameter. It outputs the distribution of abundances, gas and dust temperatures, level populations and emissivities of the most important coolants. In the present models we use 33 species (including e$^-$) and 330 reactions from the UMIST2012 network \citep{McEl13}.



We use a suite of one-dimensional uniform slabs with densities of $n_{\rm H}=10^{2,3,4,5}$\,cm$^{-3}$, interacting with four CRIR of $\zeta_{\mathrm{cr}}=(5,10,50,100)\times$10$^{-17}\,{\rm s}^{-1}$.
In addition to that and inspired from hydrodynamical simulations which indicate the existence of a connection between the local (effective) $A_{\rm V}$ and the local total number density, we use a `variable density slab' which represents the $A_{\rm V}$-$n_{\rm H}$ relation discussed in \citet{Bisbas2019} and \citet{Gaches2022}. This slab covers densities from $10^2$ to $10^{5.5}\,{\rm cm}^{-3}$ and is found to estimate reasonably well the average local PDR conditions of three-dimensional clouds (Bisbas et al., \textit{in prep.}). It is thus interesting to explore whether it can reproduce the observed abundances of species. Finally, we also include the models $\mathscr{L}$ and $\mathscr{H}$ of \citet{Padovani2018} that account for CR attenuation as a function of the local column density and using the treatment of \citet{Gaches2022}. Model $\mathscr{L}$ corresponds to a ``low'' cosmic-ray spectrum based on Voyager-1 data whereas $\mathscr{H}$ to a ``high'' spectrum \citep[see][for details]{Padovani2018}.

The FUV intensity at all times is taken to be $\chi/\chi_0$ = 100, normalized to the spectral shape of \citet{Draine1978} (see Section \ref{sec:uv}). The latter is considered as a plane-parallel field that impinges from one direction. The diffuse component of FUV radiation due to dust scattering is neglected in this work.
In our network, we include the evolution of relevant species such as H$_2$, carbon-bearing (e.g., CH, CO), and oxygen-bearing (e.g., OH, HCO$^+$) molecules.
For the constant density slabs, the maximum value of $A_{\rm V}$ is taken to be 20\,mag. 

\begin{table}[!hbpt]
\begin{tabular}{cc}
\hline
\hline
Element  & Abundance \\
\hline
C$^+$  &  2$\times$10$^{-4}$    \\
He  &  1$\times$10$^{-1}$   \\
O  &  3$\times$10$^{-4}$   \\
H$_2$  &  3$\times$10$^{-1}$ \\
H  &  4$\times$10$^{-1}$ \\
\hline
\end{tabular}
\caption{Initial abundances used in the chemical model.}
\label{tab:abundance}
\end{table}


\begin{figure*}
\includegraphics[width=1.0\linewidth]{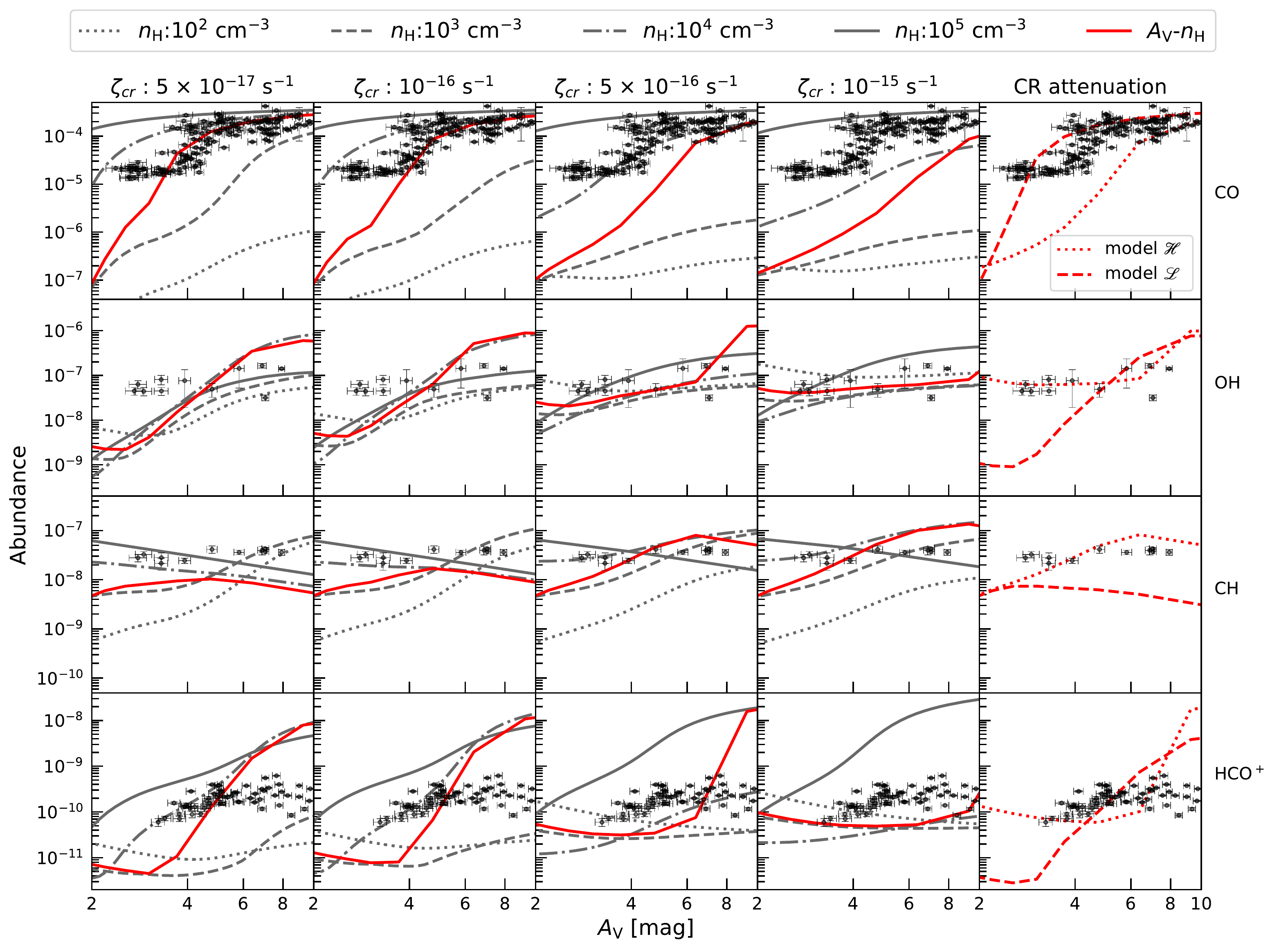}
\caption{From top-to-bottom: the abundance of CO, OH, CH, and HCO$^+$ versus $A_{\rm V}$. The scattered black dots represent the observations. Lines curves represent the model results. The four different densities are shown in different line styles (dotted line is for $n_{\rm H}=10^2\,{\rm cm}^{-3}$, dashed line for $10^3$, dot-dashed line for $10^4$ and solid line for $10^5\,{\rm cm}^{-3}$). Red lines denote the model with the variable density distribution resulting from the $A_{\rm V}$-$n_{\rm H}$ relation of \citet{Bisbas2019}. The left four columns correspond to different CRIR values used in the model ($\zeta_{\rm CR}=5,10,50,100\times10^{-17}\,{\rm s}^{-1}$). The rightmost column shows the $A_{\rm V}$-$n_{\rm H}$ model in which the CR attenuation models $\mathscr{H}$ and $\mathscr{L}$ of \citet{Padovani2018} are considered, respectively.}
\label{fig:1dpdr}
\end{figure*}

\subsection{Modelling and chemistry of CO}\label{sec:co model}

 Figure~\ref{fig:1dpdr} illustrates the results of all our calculation. Overall, changing the input environmental parameters results in a significant spread in the calculate abundances, as can be seen.
More specifically, at a given CRIR value the resultant abundance of CO (top row) increases with increasing the density and extinction (2 $<$ $A_{\rm V}$ $<$ 10\,mag). This is because the photodissociation process dominates the destruction of CO. CO still builds up its abundance at low extinction regions, while it maximizes at high extinction regions.

CO can form either through the OH channel \citep{van1988, Bisbas2017} following the reactions:
\begin{equation}
\begin{matrix}
\rm C^+ + OH \rightarrow CO^+ +H,\\
\rm CO^+ + H_2 \rightarrow HCO^+ + H,\\
\rm HCO^+ + e \rightarrow CO + H,
\end{matrix}
\label{eq:oh2co}
\end{equation}
or via the CH channel \citep{van1988} following the reaction:
\begin{equation}
\rm CH + O \rightarrow CO + H.
\label{eq:ch2co}
\end{equation}
In low density models ($n_{\rm H}$ = 10$^2$, 10$^3$\,cm$^{-3}$), the formation of CO is dominated by OH channels (see Section \ref{sec:alpha}). Since the abundance of OH is proportional to the CRIR \citep{Bialy2015}, higher CRIR will promote the formation of OH as well as CO (reaction \ref{eq:oh2co}). Thus, the resultant abundance of CO will increase with the increasing CRIR at low $A_{\rm V}$ ($A_{\rm V}$ $<$ 4\,mag) regions, as shown in Figure \ref{fig:1dpdr}.

However, in regions of high extinctions and where molecules are well-shielded from UV photons, CO is mainly destroyed through the reactions \citep{Dalgarno2006, Indriolo2012, Bisbas2015, Bisbas2021}:
\begin{equation}
\rm He^+ + CO \rightarrow O + C^+ + He,
\label{eq:heco}
\end{equation}
and
\begin{equation}
\rm H^+_3 + CO \rightarrow H_2 + HCO^+,
\label{eq:co2hcop}
\end{equation}
where the abundances of He$^+$ and H$^+_3$ depend on the value of CRIR. Thus, the abundance of CO decreases as CRIR increases at high $A_{\rm V}$ ($A_{\rm V}$ $>$ 4\,mag) regions. In the dense cloud model (10$^{5}$\,cm$^{-2}$), the CRIR has little effect on the predicted CO abundance, the latter being in the range of $(1-2)\times10^{-4}$.

At a density of $n_{\rm H}$ = 10$^4$\,cm$^{-3}$, CO decreases by an order of magnitude when the CRIR increases from 5$\times$10$^{-17}$\,s$^{-1}$ to 10$^{-15}$\,s$^{-1}$, which implies that the destruction of CO from He$^+$ is important in addition to the photodissociation. We find that the best model with constant $n_{\rm H}$ and $\zeta_{\rm cr}$ that approaches the observed CO abundances is for $n_{\rm H}$ = 10$^4$\,cm$^{-3}$ and $\zeta_{\rm cr}$ = 5$\times$10$^{-16}$\,s$^{-1}$. 

The predicted abundance of the $A_{\rm V}$-$n_{\rm H}$ model falls in between constant density models with $n_{\rm H}$ of 10$^3$ and 10$^4$\,cm$^{-3}$. The effect of CR attenuation model $\mathscr{H}$ is similar to the one with $\zeta_{\rm cr}$ = 5$\times$10$^{-16}$\,s$^{-1}$, while model $\mathscr{S}$ to the one with $\zeta_{\rm cr}$ = 5$\times$10$^{-17}$\,s$^{-1}$. We find that the $A_{\rm V}$-$n_{\rm H}$ simulations with high CRIR ($\zeta_{\rm cr}$ $\geq$5$\times$10$^{-16}$\,s$^{-1}$ or model $\mathscr{H}$) underestimates the observed abundance of CO at low $A_{\rm V}$. Thus, if the gas density follows a similar distribution as the $A_{\rm V}$-$n_{\rm H}$, the CRIR should be between the CR attenuation models $\mathscr{H}$ and $\mathscr{L}$.

\subsection{Modelling and chemistry of OH}\label{sec:oh model}

In the second row of Fig.~\ref{fig:1dpdr}, the abundance of OH is plotted. Overall, the abundance of this species increases as $A_{\rm V}$ increases from 2 to 10\,mag for a given CRIR value, except for the low density model ($n_{\rm H}$ = 10$^2$\,cm$^{-3}$) at high CRIR ($\zeta_{\rm cr}$ $\geq$5$\times$10$^{-16}$\,s$^{-1}$). The formation of OH starts from the proton transfer reaction \citep{van1986,Bialy2015,Bisbas2017}: 
\begin{equation}
\rm H^+_3 + O \rightarrow OH^+ + H_2,
\end{equation}
or through the charge transfer reaction:
\begin{equation}
\rm H_2 + O^+ \rightarrow OH^+ + H.
\end{equation}
OH$^+$ is then hydrogenated to form H$_2$O$^+$ and H$_3$O$^+$, which leads to the rapid formation of OH through the recombination reaction with electrons \citep{Bialy2015}. The destruction of OH is dominated by photodissociation (at low $A_{\rm V}$) and abundant species (e.g., C$^+$, C, O).

A larger flux of low-energy CRs produce more H$^+_3$ and H$^+$, which reacts with oxygen atoms to produce more OH$^+$. Numerically, the abundance of OH is proportional to the CRIR at low $A_{\rm V}$ (Luo et al. in prep). This is particularly noticeable in Figure \ref{fig:1dpdr} at low $A_{\rm V}$, where the abundance of OH increases with the increasing CRIR. The increasing abundance of OH can reasonably explain the increasing trend of CO and HCO$^+$ abundances with higher CRIRs at low density and low $A_{\rm V}$.

At high $A_{\rm V}$, higher CRIR will increase the abundance of C$^+$, C, and electrons, which could accelerate the destruction of OH \citep{Bialy2015}.
The development of OH abundance versus $A_{\rm V}$ tends to be flatten (saturates) as CRIR increases, independent on the assumed density. 
From our model prediction, it appears that the formation of OH is preferred for $n_\mathrm{H}$ = 10$^4$\,cm$^{-3}$ at low CRIR ($\zeta_{\mathrm{cr}} \leq $10$^{-16}$\,s$^{-1}$) and high $A_{\rm V}$ ($>$4\,mag) environments, in which the abundance of OH reaches a maximum among these models. 

Among all these simulations, models with low CRIR values ($\zeta_{\rm cr}$ $\leq$10$^{-16}$\,s$^{-1}$) would always underestimate the observed OH abundance at low $A_{\rm V}$ (by a factor of few to an order of magnitude). A higher CRIR model ($\zeta_{\mathrm{cr}} \geq$ 5$\times$10$^{-16}$\,s$^{-1}$) is needed to explain the observed abundance at low $A_{\rm V}$. The variable density slab model with both high CRIR or CR attenuation model $\mathscr{H}$ can reasonably reproduce the observed abundance of OH.

\subsection{Modelling and chemistry of CH}\label{sec:ch model}

The third row of Fig.~\ref{fig:1dpdr} illustrates the results for the CH abundance. CH is an alternative channel to form CO through Reaction \ref{eq:ch2co}. Overall, the calculated abundance of CH increases with increasing the density at low $A_{\rm V}$ ($\leq$4\,mag). The formation of CH is starting from C$^+$ and H$_2$ through a sequence of reactions \citep{Bialy2015,Bisbas2019}:
\begin{equation}
\begin{matrix}
\rm C^+ + H_2 \rightarrow CH^+_2 + h\nu,\\
\rm CH^+_2 +e \rightarrow CH + H,\\
\rm CH^+_2 + H_2 \rightarrow CH^+_3 + H,\\
\rm CH^+_3 + e \rightarrow CH + H_2,
\end{matrix}
\label{eq:cp2ch}
\end{equation}
and the destruction of CH is dominated by photodissociation at low extinction ($A_{\rm V}$ $\leq$4\,mag) regions. The estimated abundance of this species increases monotonically with increasing $A_{\rm V}$ for the low density models ($n_\mathrm{H}$ = 10$^2$, 10$^3$\,cm$^{-3}$), which is insensitive to CRIR at low $A_{\rm V}$ ($\leq$4\,mag).

However, at high extinction regions where the electron fraction is low, the formation of CH through reactions \ref{eq:cp2ch} are not efficient. The formation of CH would involve reactions \citep{Bialy2015}:
\begin{equation}
\begin{matrix}
\rm C + H^+_3 \rightarrow CH^+ + H_2,\\
\rm CH^+ + H_2 \rightarrow CH_2^+ + H,\\
\rm CH_2^+ + H_2 \rightarrow CH_3^+ + H,\\
\rm CH^+_3 + e \rightarrow CH_2 + H,\\
\rm CH_2 + H \rightarrow CH + H_2,
\end{matrix}
\end{equation}
and the destruction of CH is mainly contributed by abundant neutral species (e.g., H, O) through the following reactions \citep{Bialy2015}:
\begin{equation}
\begin{matrix}
\rm CH + H \rightarrow C + H_2,\\
\rm CH + O \rightarrow CO + H.
\end{matrix}
\end{equation}
Thus, the increasing CRIR may promote the formation of CH at high extinction regions. As seen in Figure \ref{fig:1dpdr}, the CH abundance slightly increases with the increasing CRIR at high $A_{\rm V}$ ($\geq$4\,mag). The abundance of CH increases with the increasing CRIR for high density models, while it decreases for low density models. This is possibly due to the different dominant destruction pathways. For low density models, the increasing CRIR would increase the destruction of CH by H, while the destruction of CH at high density is dominated by O.

At low CRIR values, a single density model cannot reproduce the observed CH abundance. The latter should most likely exist in an environment with higher CRIR ($\zeta_{\mathrm{cr}} \geq$ 5$\times$10$^{-16}$\,s$^{-1}$) and with moderate densities ($n_\mathrm{H}$ = 10$^3$ -- 10$^4$\,cm$^{-3}$). 
In regards to the variable density slab model, the best-fit CRIRs includes $\zeta_{\rm cr}=5\times10^{-16}\,{\rm s}^{-1}$ and the CR attenuation model $\mathscr{H}$. Lower CRIRs or CR the attenuation model $\mathscr{L}$ appear to underestimate the CH abundance at all times.

\subsection{Modelling and chemistry of HCO$^+$}\label{sec:hcop model}

In the bottom row of Fig.~\ref{fig:1dpdr}, the corresponding abundances of HCO$^+$ are illustrated. The abundance of HCO$^+$ varies in a more complex way for different model parameters than the above abundances. In general, the low CRIR models  ($\zeta_{\mathrm{cr}} \leq$ 10$^{-16}$\,s$^{-1}$) result in an HCO$^+$ abundance that increases with increasing $A_{\rm V}$. High density models ($n_\mathrm{H}$ = 10$^4$, 10$^5$\,cm$^{-3}$) estimate a much higher (up to two orders of magnitude) HCO$^+$ abundance than the corresponding ones for low densities. For the high CRIR cases ($\zeta_{\mathrm{cr}} \geq$ 10$^{-16}$\,s$^{-1}$), only the highest density model ($n_\mathrm{H}$ = 10$^5$\,cm$^{-3}$) remains unchanged, while the rest ($n_\mathrm{H}=10^2-10^4\,{\rm cm}^{-3}$) result in a similar abundance of HCO$^+$ in the range $(5-30)\times10^{-11}$.

Different from the above three species whose destruction pathways are controlled by both photodissociation and abundant species (e.g., H, O, C$^+$), the destruction of HCO$^+$ is always done by electron recombination reaction:
\begin{equation}
\rm HCO^+ + e \rightarrow CO + H.
\label{eq:hcop2co}
\end{equation}
At low $A_{\rm V}$, the photoionization of carbon contributes most of the free electrons. At high extinction regions, free electrons are controlled by CR ionization of H$_2$ and He.
The formation of HCO$^+$ at low density is through OH channels (Reaction \ref{eq:oh2co}), in which HCO$^+$ is the precursor of CO. Thus, the abundance of HCO$^+$ increases with the increasing CRIR, which is similar to that of OH. However, at high density, CO will be the precursor of HCO$^+$ through Reaction \ref{eq:co2hcop}. Thus, with the increasing CRIR, the competing processes of the formation Reactions \ref{eq:oh2co}, \ref{eq:co2hcop} and destruction reaction \ref{eq:hcop2co} control the variance of HCO$^+$ abundance.

Since the optical depth of HCO$^+$ could be quite large at high $A_{\rm V}$ (see Section \ref{sec:uncertainties}), the abundance of HCO$^+$ at high $A_{\rm V}$ is less reliable and hard to justify. Most of the observed HCO$^+$ abundance are in agreement with a moderate density model ($n_{\rm H}$ $\sim$ 10$^4$\,cm$^{-3}$). The variable density slab models with low CRIR ($\zeta_{\rm cr}$ $\leq$10$^{-16}$\,s$^{-1}$) underestimate the abundance of HCO$^+$ at low $A_{\rm V}$ by an order of magnitude, which may suggest a CRIR higher than 10$^{-16}$\,s$^{-1}$.


\section{Discussion}\label{sec:discussion}

\subsection{Uncertainties}\label{sec:uncertainties}

The observational uncertainties and error propagation are considered in the above calculations. However, the prerequisite of assumptions (e.g., excitation temperature, optically thin) in deriving the relative abundances of CO, OH, CH, and HCO$^+$ to H$_2$ may bring additional uncertainties, which could alter the final results in some cases. Here, we describe the uncertainty analysis, which helps us understand to what extent the prerequisite of assumptions can influence the final results.



Since H$_2$ cannot be measured directly in the cold gas, to obtain the column density of H$_2$ ($N_{\rm H_2}$), we have deducted the H\,{\sc i} column density from the total gas column density. The typical conversion factor ($N_\mathrm{H}$/$A_{\rm V}$, see section \ref{sec:abundance}) has been widely used to convert extinction to the total gas column density. However, different $N_\mathrm{H}$/$A_{\rm V}$ ratios have been found in different environments by recent observations. For example, in diffuse LOSs the $N_\mathrm{H}$/$A_{\rm V}$ ratios could be up to 2 times higher than the typical one \citep[][]{Liszt2014,Nguyen2018}. X-ray observations found that the ratio is 50\% higher than the typical one in SNRs \citep[][]{Guver2009,Foight2016}. In our case, IC~348 is part of the GMC, which lies far enough from the nearby SNRs. Considering that the $A_{\rm V}$ value is in the range of $2-10$\,mag, we thus adopt the value of \citet{Bohlin1978} (see \S\ref{sec:abundance}).

\begin{figure}
\includegraphics[width=1.0\linewidth]{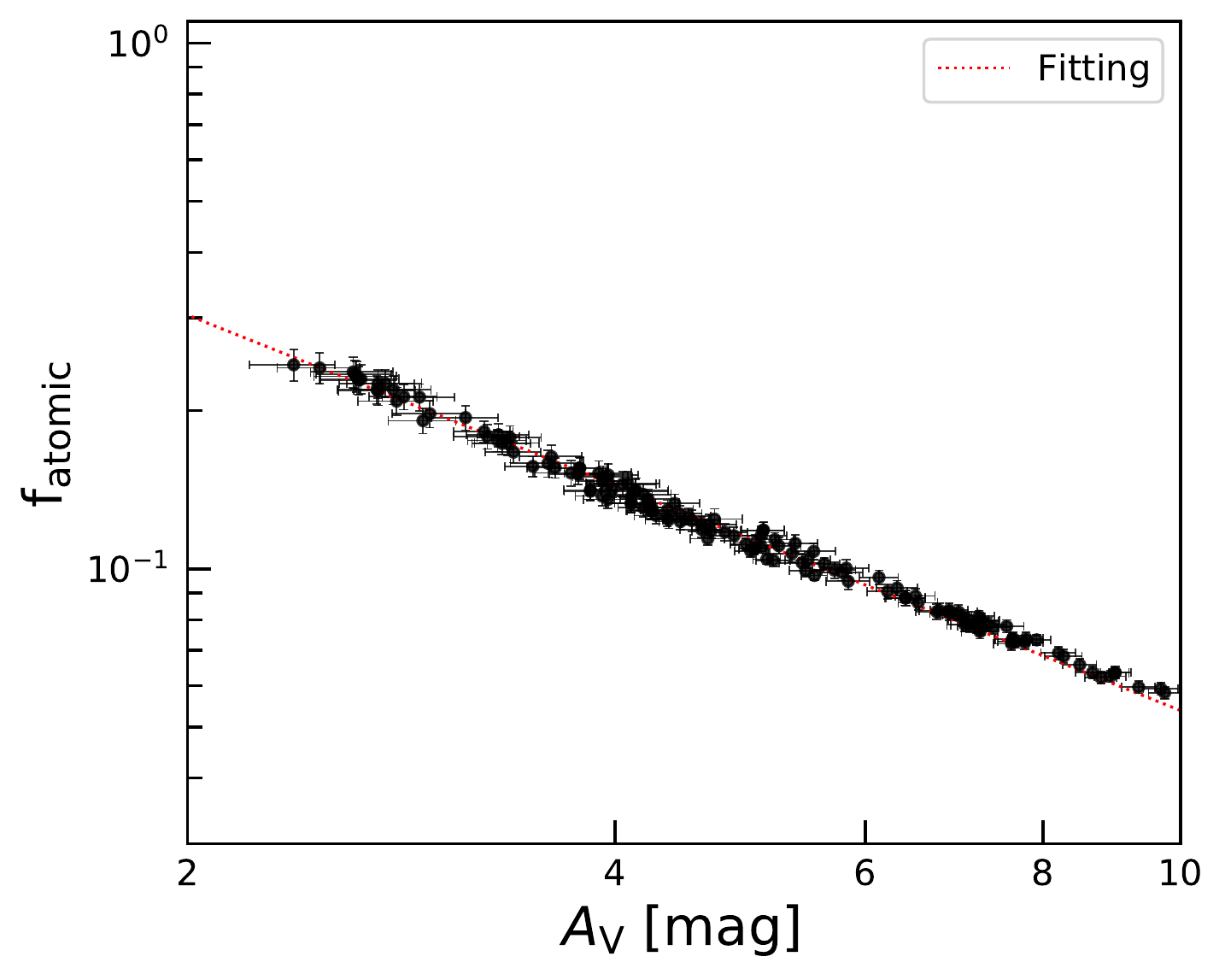}
\caption{Atomic fraction ($f_\mathrm{atomic}$) versus $A_{\rm V}$. The atomic fraction decreases with increasing $A_{\rm V}$. The maximum value of $f_\mathrm{atomic}$ is $\sim$24\% at 2.4\,mag, and the minimum is $\sim$6\% at 9.7\,mag.}
\label{fig:f_atomic}
\end{figure}

In our calculation, we assume that the H\,{\sc i} gas is optically thin. 
We note that the H\,{\sc i} column density would be underestimated by $\sim$10\% with an optical depth correction from LOSs against background quasars \citep{Lee2015}. To quantify the uncertainties of $N_{\rm H_2}$ induced by the corresponding ones from $N_\mathrm{H\,{\sc I}}$, we have calculated the atomic fraction in this cloud, defined as $f_\mathrm{atomic}$ = $N_{\rm H\,{\sc I}}$/$N_{H}$. The atomic fraction decreases with  increasing $A_{\rm V}$, which follows a power law: $f_\mathrm{atomic}$ = (0.64 $\pm$ 0.01) $\times$ ($\frac{A_{\rm V}}{[\mathrm{mag}]}$)$^{(-1\pm0.01)}$ in the $A_{\rm V}$ range of 2$\sim$10\,mag (as shown in Figure \ref{fig:f_atomic}). The value of $f_\mathrm{atomic}$ ranges from 6\% to 24\%, with a mean value of 12\%. Therefore, the molecular gas dominates the gas components in IC~348, and the uncertainties of $N_{\rm H_2}$ induced by H\,{\sc i} would be in the range of 0.6\% to 3\%. Thus, it would not alter the final results of relative abundance. 

The excitation temperature of OH was first measured from absorption against quasars, which is in the range between $4-8\,{\rm K}$ \citep{Dickey1981}. \citet{Liszt1996} measured the excitation temperature of OH through both emission and absorption profiles toward 8 quasars in which they considered a typical value of $T_\mathrm{ex} - T_\mathrm{CMB}$ = 1\,K. Recent work from an even larger sample of quasar-absorption systems ($\sim$44) by \citet{Li2018} found that the majority ($\sim$90\%) of OH components have an excitation temperature within 2\,K of the Galactic emission background. In addition, this value can be generalized to various molecular cloud environments \citep[e.g., dark clouds, $Planck$ Galactic Cold Clumps (PGCCs),][]{Tang2021a}. However, due to lack of constraint on the excitation condition, the uncertainty of the empirical correlation of $T_\mathrm{ex}$ - $T_\mathrm{bg}$ = 2\,K could be up to an order of magnitude when $T_\mathrm{ex}$ is close to $T_\mathrm{bg}$. Following the $T_\mathrm{ex}$ - $T_\mathrm{bg}$ = 2\,K \citep{Li2018}, the maximum optical depth of OH can be obtained by solving the radiative transfer Eqn.~\ref{eq:t_mb}, which is $\sim$0.7 at the peak intensity ($\sim$1\,K toward p03). Thus, the uncertainty of OH column density as well as the abundance induced by optical depth could be underestimated by 38\%.

\begin{figure}
\includegraphics[width=1.0\linewidth]{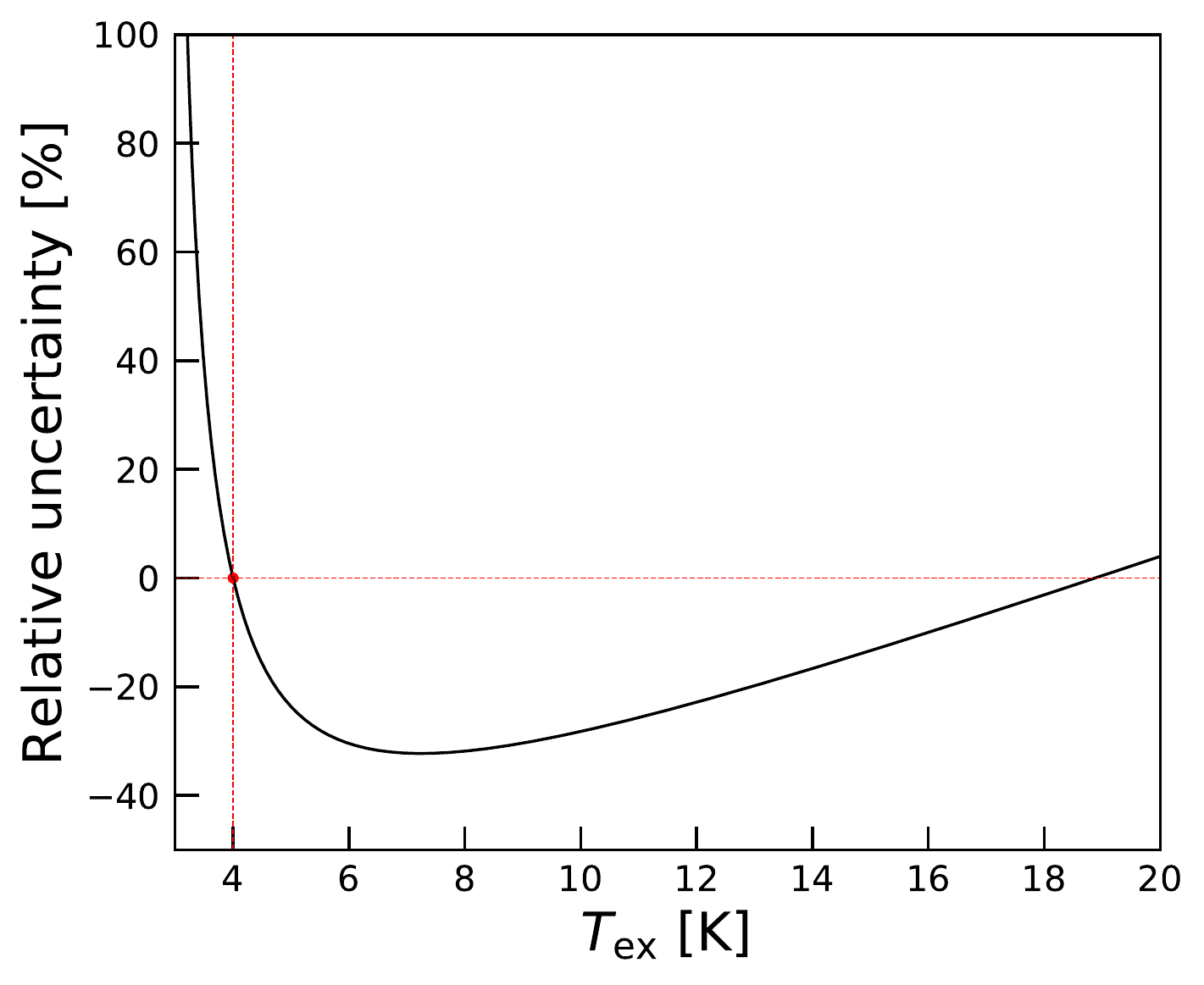}
\caption{The dependence of relative uncertainty of the HCO$^+$ column density on the excitation temperature. The intersection point (red dot) of the dashed lines represent the reference $T_\mathrm{ex}$ (4\,K) adopted in this work.}
\label{fig:hcop_err}
\end{figure}

The calculation of the CH column density is based on the common assumption that the excitation temperature of CH is much higher than the background temperature. However, combined observations of optical/UV absorption and radio emission \citep{Dailey2020}, ON--OFF observations against quasars using Arecibo \citep{Tang2021b}, and interferometry observations against bright galactic continuum sources \citep{Jacob2021} have found that the $T_\mathrm{{ex}}$ values have a large scatter toward different LOSs, which could deviate from the canonical value. We should note that in some extreme cases when the $T_\mathrm{{ex}}$ reaches a value of -0.3\,K, the derived column density of CH would decrease by a factor of $\sim$10.

The excitation temperature assumption of HCO$^+$ could be biased by the volume density of the cloud. In high density regions, HCO$^+$ could be thermally excited, while in low density regions it can be sub-thermally excited. If we consider the differences of column density induced by the excitation temperature, in the case of $T_\mathrm{ex}$ varying from 4--20\,K, the relative uncertainty (with respect to the value at $T_\mathrm{ex}$ = 4\,K) could be varied from -32\% to 5\% (Figure \ref{fig:hcop_err}). 

We also note that a very low value of $T_\mathrm{{ex}}$ (e.g., 0.1\,K above the CMB) is frequently observed in diffuse clouds \citep[$n_\mathrm{H}$ $\ll$ 10$^3$\,cm$^{-2}$,][]{Muller2009, Godard2010, Luo2020}. As can be seen in Figure~\ref{fig:hcop_err}, the relative uncertainty of HCO$^+$ column density has a steep rise at $T_\mathrm{ex}$ $<$4\,K. If we adopt $T_\mathrm{{ex}}$ - $T_\mathrm{{CMB}}$ = 0.1\,K, the column density of HCO$^+$ would increase by over an order of magnitude. However, such extreme case would not happen in the studied cloud, especially for the ICRs. Otherwise, the brightness temperature of HCO$^+$ ($\sim$10\,mK) would have been far below the sensitivity limit (1\,$\sigma$ $\sim$20\,mK). 

The column density of HCO$^+$ would be underestimated for high extinction regions where the emission of HCO$^+$ is optically thick. \citet{Baras2021} found that the $^{12}$C/$^{13}$C isotope ratio inferred from H$^{12}$CO$^+$/H$^{13}$CO$^+$ in high extinction regions ($A_{\rm V}$ $>$ 5\,mag) is much lower than the average value of the solar neighborhood. Thus, the optically thin isotope line (e.g., H$^{13}$CO$^+$, $J$=1--0) is needed to constrain the column density of HCO$^+$ in high extinction regions.

\subsection{The fraction of CO formation through OH channels}\label{sec:alpha}

\begin{figure}
\includegraphics[width=1.0\linewidth]{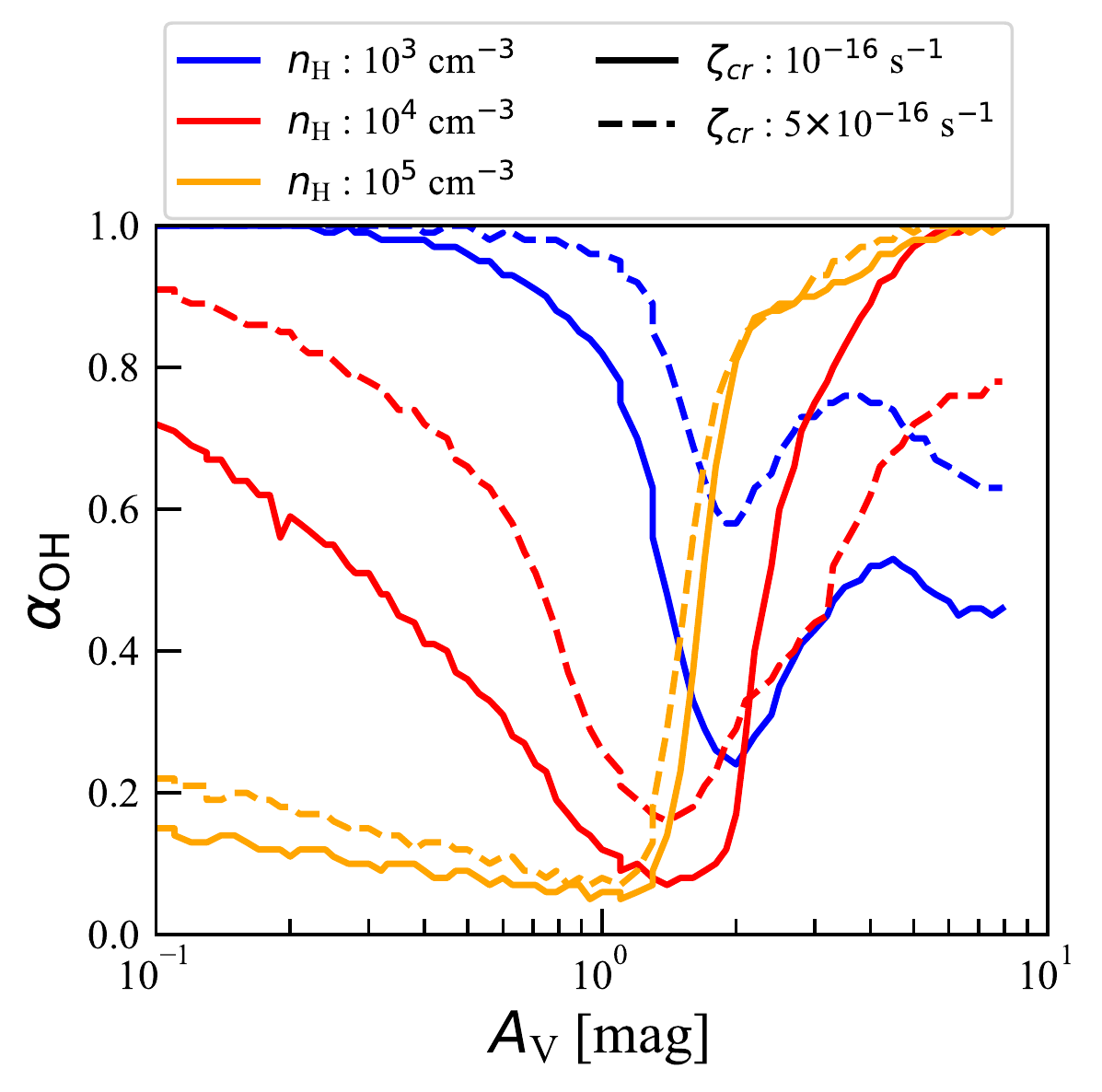}
\caption{The variances of $\alpha_\mathrm{OH}$ ratio against the visual extinction and for different PDR models. $\alpha_\mathrm{OH}$ represents the fraction of CO formation through OH channel against the total CO formation. Colors represent different densities (blue for $10^3$, red for $10^4$ and yellow for $10^5\,{\rm cm}^{-3}$) and line styles represent two CRIR values (solid for $\zeta_{\rm CR}=10^{-16}$ and dashed for $5\times10^{-16}\,{\rm s}^{-1}$). $\alpha_\mathrm{OH}$ decreases with increasing $A_{\rm V}$ and $n_{\rm H}$ in low extinction regions, while it increases with increasing CRIR.}
\label{fig:roh}
\end{figure}

As mentioned in Section \ref{sec:co model} CO can form through either the OH channel (Reaction \ref{eq:oh2co}) or the CH channel (Reaction \ref{eq:ch2co}). The variance of the ISM environmental parameters can largely influence the abundance of OH and CH, further affecting the formation of CO. We now define with $\alpha_{\rm OH}$ the fraction of CO formation through the OH channel to the total CO formation. Figure~\ref{fig:roh} shows the dependence of $\alpha_\mathrm{OH}$ versus the $A_{\rm V}$, at different densities and CRIR values. 

%
%
The value of $\alpha_\mathrm{OH}$ decreases with increasing $A_{\rm V}$ and $n_{\rm H}$ at low $A_{\rm V}$, indicating less CO formation through OH. Instead, in this regime CO forms via the CH channel. When the extinction increases further, $\alpha_\mathrm{OH}$ increases. All models (except for $n_\mathrm{H}$ = 10$^4$\,cm$^{-3}$ at high extinction) exhibit higher $\alpha_\mathrm{OH}$ at high CRIR, indicating that more CO forms via OH with increasing CRs. This is because the abundance of CH is almost independent of CRIR at low $A_{\rm V}$, while the abundance of OH increases as CRIR increases. Overall, at high densities and high extinctions (e.g., internal of molecular cloud), CO is mainly formed through the OH channel. In the low-to-intermediate extinction regions, the CH channel may contribute most of the CO formation (up to 95\%). In the case of IC~348 and for the estimated gas density and CRIR from our models ($n_{\rm H}$ = 10$^4$\,cm$^{-3}$, $\zeta_\mathrm{cr}$ = 5$\times$10$^{-16}$\,s$^{-1}$), $\alpha_\mathrm{OH}$ is in the range of $30-80\%$.



\subsection{The CRIR in IC~348}\label{sec:CRIR}

By comparing the observational results with our model predictions in Section \ref{sec:model}, we find that the one-dimensional slab PDR model underestimates the abundances of OH and CH in the lower extinction regions ($A_{\rm V} <$ 4\,mag) if we adopt a low CRIR value ($\zeta_{\rm cr}$ = 5$\times$10$^{-17}$ or 10$^{-16}$\,s$^{-1}$). For the constant $n_{\rm H}$ and $\zeta_{\rm cr}$ models, the observations mostly coincide with $n_{\rm H}$ = 10$^4$\,cm$^{-3}$ and $\zeta_{\mathrm{cr}}=5\times10^{-16}\,{\rm s}^{-1}$. The variable density models using the  CR attenuation model $\mathscr{H}$ can best reproduce the abundances of OH, CH, and HCO$^+$ (except for the large uncertainty of the observed HCO$^+$ abundance at high $A_{\rm V}$, see Section \ref{sec:hcop model} and \ref{sec:uncertainties}), while the model underestimates the CO abundance at low $A_{\rm V}$. The variable density models with the CR attenuation model $\mathscr{L}$ can reasonable explain the abundance of CO, while it underestimates the abundances of other species. As we do not expect for a single model to explain all observational data, our results indicate that the density distribution of the star-forming cloud may also differ from the $A_{\rm V}$-$n_{\rm H}$ relation, but not significantly.

To place more stringent constraints on $\zeta_{\rm cr}$ and $n_{\rm H}$ at different positions, we have run PDR models with $\zeta_{\rm cr}$ from 10$^{-17}$ to 10$^{-14}$\,s$^{-1}$ and $n_{\rm H}$ from  10$^2$ to 10$^{5.5}$\,cm$^{-3}$. We vary $\zeta_{\rm cr}$ and $n_{\rm H}$ to find the optimum model by minimizing the reduced $\chi^2$ function:
\begin{equation}
\rm \chi^2 = \frac{1}{N}\sum^N_{i=1}  \frac{\left ( f_{obs}^i - f_{model}^i \right )^2}{{\sigma^i_{obs}}^2},
\end{equation}
where $\rm {f_{obs}^i}$ and ${\rm \sigma^i_{obs}}$ are the molecular abundance and uncertainty of the $i$th species from observations, respectively. $\rm {f_{model}^i}$ is the abundance from PDR models and N is the number of the species. 

\begin{figure*}
\includegraphics[width=1.0\linewidth]{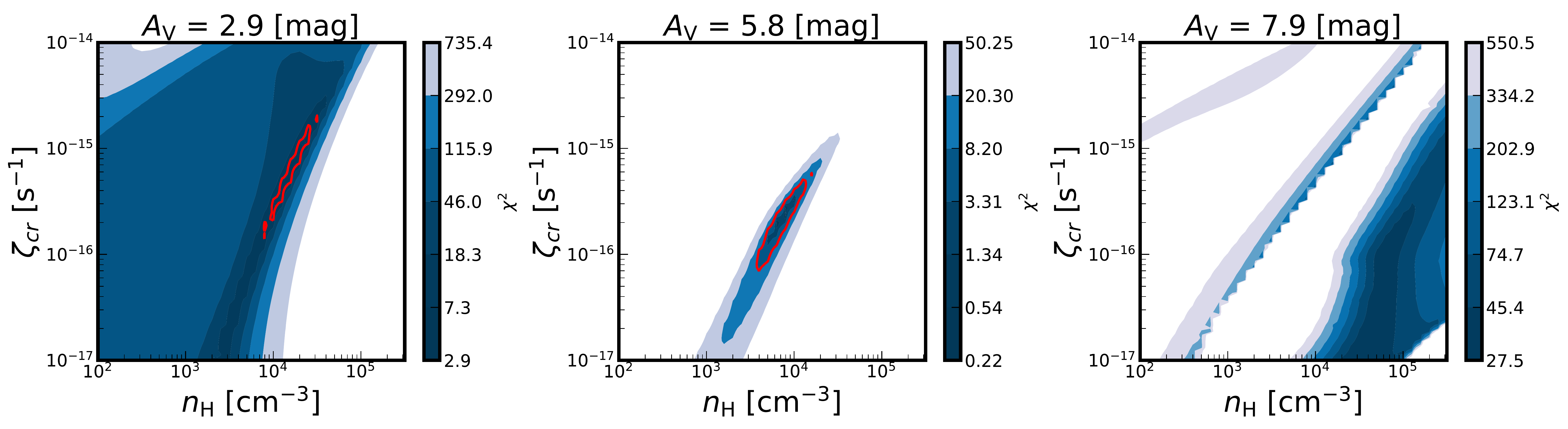}
\caption{The $\chi^2$ distribution at three representative positions in the given parameter space. The $A_{\rm V}$ values are given at the top of each panel. Red contours represent the boundary of where the deviation between modelled values and observations is 3$\sigma$.}
\label{fig:chi2}
\end{figure*}

Figure \ref{fig:chi2} shows the reduced $\chi^2$ distributions at low (left, $A_{\rm V}$ = 2.9$\pm$0.2\,mag), intermediate (middle, $A_{\rm V}$ = 5.8$\pm$0.2\,mag), and high extinction positions (right, $A_{\rm V}$ = 7.9$\pm$0.2\,mag) in the above parameter space. The higher values of $\chi^2$ are colored with bright blue and lower colored with dark blue. Red contours represent the boundary of the best-fit parameter where the deviation between modeled values and observations is 3$\sigma$. We stress here that the reduced $\chi^2$ fitting result is the optimum solution numerically rather than physically. In some cases, the fitting results will appear over-fit ($\chi^2$ $<$1), which does not mean the best-fit parameters is physically better than those parameters that produce a larger $\chi^2$ value. In a few cases, the minimum $\chi^2$ value is larger than the 3$\sigma$ contour (e.g., right panel of Figure \ref{fig:chi2}). This is because the uncertainty of the observational values are so small that even a small deviation between models and observations would lead to a large $\chi^2$ value. Moreover, the minimal $\chi^2$ value is located in the boundary of our parameter space. Thus, only upper/lower limits are given.

From low to high extinction (left to right panel of Fig.~\ref{fig:chi2}), the best-fit results of $n_{\rm H}$ and $\zeta_{\rm cr}$ are (1.6$^{+0.4}_{-0.3}$,  0.8$^{+0.2}_{-0.2}$, 4.5$^{+0.1}_{-0.1}$) $\times$10$^4$\,cm$^{-3}$ and (6.6$^{+3.4}_{-2.3}$, 2.7$^{+0.6}_{-0.8}$, $\leq$0.1) $\times$10$^{-16}$\,s$^{-1}$, respectively. The fitting results toward 10 positions (2 positions without OH or CH detections are excluded) are summarised in Table \ref{tab:chi2}. 
The average value of CRIR at low extinction regions ($A_{\rm V}$ $<$4\,mag) is (4.7$\pm$1.5)$\times$10$^{-16}$\,s$^{-1}$. This value is consistent with measurements of H$^+_3$ towards two nearby massive stars ($\zeta_{\rm cr}$ = 5.55$\pm$3.18$\times$10$^{-16}$\,s$^{-1}$ toward $\zeta$ Per) in the Per OB 2 Association. A notable trend from low $A_{\rm V}$ to high $A_{\rm V}$ is that the CRIR decreases from 6.6$^{+3.4}_{-2.3}$ $\times$10$^{-16}$\,s$^{-1}$ to $\leq$2$\times$10$^{-16}$\,s$^{-1}$. The decreasing trend of CRIR from low to high $A_{\rm V}$ in IC~348 seems to support the hypothesis of $\zeta_{\rm cr}$ gradient as a function of $A_{\rm V}$, as predicted by theoretical models \citep{Padovani2018, Padovani2022, Gaches2022}. 

We note that Position IDs 8, 10 and ID 9 have similar $A_{\rm V}$ but significant $\zeta_{\rm cr}$ (an order of magnitude difference). The most possible reason is that the $n_{\rm H}$ value from the $\chi^2$ fitting has been underestimated. As shown in Table \ref{tab:chi2}, the best-fit $n_{\rm H}$ values at Position IDs 8 and 10 are more than an order of magnitude lower than that of ID 9. However, the gas density inferred from CS lines at similar $A_{\rm V}$ is above 10$^5$\,cm$^{-3}$ \citep{Baras2021}. Moreover, the bright HCO$^+$ and HCN emission ($T_{\rm mb}$ $\sim$ 1\,K) toward Position IDs 8 and 10 is unlikely originated from low density gas. Therefore, the numerical best-fit $n_{\rm H}$ value may not be physical at Position IDs 8 and 10. If we limit the gas density to a higher (and more reasonable) value, the best-fit $\zeta_{\rm cr}$ value would move to a higher value (as can be expected from the right panel of Fig.~\ref{fig:chi2}). On the other hand, the clumpy structure of the cloud may also bias the results. In the modelling we performed $\chi^2$ fitting, we have only considered a 1D cloud structure with a uniform density for each model. However, the cloud structure may be clumpy and filamentary. The results suggest that a better understanding of the cloud density and structure is helpful to constrain the CRIR, especially at high $A_{\rm V}$ positions.

\begin{table}
\begin{tabular}{cccc}
\hline
\hline
Position  & $A_{\rm V}$  &  best-fit parameters & \\
\cline{2-4}
ID& [mag]  & log$_{10}$ ($n_{\rm H}$/[cm$^{-3}$]) &  log$_{10}$ ($\zeta_{\rm cr}$/[s$^{-1}$])\\
\hline
1&2.8$\pm$0.2  &  4.1$^{+0.1}_{-0.1}$  & -15.3$^{+0.2}_{-0.2}$  \\
2&2.9$\pm$0.2  &  4.2$^{+0.1}_{-0.1}$  & -15.2$^{+0.2}_{-0.2}$   \\
3&3.3$\pm$0.2  &  3.9$^{+0.2}_{-0.1}$  & -15.5$^{+0.3}_{-0.3}$   \\
4&3.3$\pm$0.2  &  4.1$^{+0.1}_{-0.1}$  & -15.2$^{+0.2}_{-0.2}$   \\
5&3.9$\pm$0.2  &  3.9$^{+0.1}_{-0.2}$  & -15.5$^{+0.3}_{-0.1}$   \\
6&4.8$\pm$0.2  &  4.2$^{+0.1}_{-0.1}$  & -15.8$^{+0.4}_{-0.2}$   \\
7&5.8$\pm$0.2  &  3.9$^{+0.1}_{-0.1}$  & -15.6$^{+0.1}_{-0.2}$   \\
8&6.9$\pm$0.2  &  3.2$^{+0.1}_{-0.1}$  & -17.0$^{+0.1}_{-0.1}$   \\
9&7.0$\pm$0.2  &  $\geq$5.5  & -16.0$^{+0.2}_{-0.2}$   \\
10&7.9$\pm$0.2  &  4.5$^{+0.1}_{-0.1}$  & $\leq$-17.0   \\
\hline
\end{tabular}
\caption{Best-fit parameters toward 10 positions.}
\label{tab:chi2}
\end{table}


\section{Conclusions}\label{sec:conclusion}
We have performed a multiwavelength observational study of IC~348 in the Perseus molecular cloud, with OH and CH $\Lambda$-doubling lines using the Arecibo telescope, HCO$^+$ $J$ = 1--0 lines using the Delingha 13.7-m telescope, archival data from FCRAO ($^{12}$CO, $^{13}$CO $J$ = 1--0), JCMT ($^{13}$CO $J$ = 3--2), and Arecibo (GLAFA-H\,{\sc i}). We have investigated the physical and chemical properties of gas components by comparing the observed spectral lines with PDR models using {\sc 3d-pdr} and especially the CRIR. The main conclusions are as follows:

\begin{enumerate}
    \item  The line intensities of $^{12}$CO, $^{13}$CO, HCO$^+$, and CH are stronger toward the interior of cloud regions (ICRs, p01 to p06) than in the exterior regions (ECRs, p06 to p12), where they drop significantly. The average observed visual extinction in the boundary between ICRs and ECRs is  $A_{\rm V}\simeq4\,{\rm mag}$. 
    
    \item The excitation temperature of CO derived from $^{13}$CO $J=1-0$ and $3-2$ ($T^2_\mathrm{ex}$) is on average half of that derived from $^{12}$CO ($T^1_\mathrm{ex}$). The high excitation temperature of $^{12}$CO is mainly due to the high UV radiation field ($\chi/\chi_0\sim100$) in IC~348, resulting in the high brightness temperature of $^{12}$CO at the outskirts of the molecular cloud. 

    \item  Assuming an isotope ratio of $^{12}$C/$^{13}$C$=65$, the average abundance of CO in IC~348 is 1.2$\pm$0.9 $\times$10$^{-4}$, which is similar to the observed value in nearby molecular clouds. The abundance of CO toward ECRs is more than six times lower than that of ICRs. The low abundance of CO at low $A_{\rm V}$ results from both photodissociation and CR induced destruction processes.

    \item  We find a constant abundance of CH (3.3$\pm$0.7 $\times$10$^{-8}$) at both ICRs and ECRs, which is in good agreement with the measured abundance at diffuse clouds in previous optical and radio observations. The abundance decline of CH at high extinction regions does not occur in this region (up to $A_{\rm V}$ = 8\,mag).

    \item  By comparing the molecular abundances of CO, OH, CH, and HCO$^+$ with the one-dimensional PDR models, we find that the CRIR is higher ($\zeta_{\rm cr}$ = (4.7$\pm$1.5)$\times$10$^{-16}$\,s$^{-1}$) at low $A_{\rm V}$ ($<$4\,mag) and lower ($\zeta_{\rm cr}$ $\leq$ 2$\times$10$^{-16}$\,s$^{-1}$) at high $A_{\rm V}$ ($>$4\,mag). The inferred CRIR in IC~348 is consistent with the H$^+_3$ measurements toward 2 nearby massive stars. Our results may support the hypothesis of $\zeta_{\rm cr}$ gradient as $A_{\rm V}$ as predicted by theoretical models.


\end{enumerate}

Overall, we have demonstrated the success of combining the multi-wavelength observations and PDR models to constrain the CRIR in the low-to-intermediate extinction range without the need for background massive stars. With the proposed capability of high-sensitivity radio telescopes (e.g., FAST, SKA), it is possible to constrain the CRIR through the ubiquitous simple hydrides (e.g., OH, CH) in both the Milky Way and external galaxies. Despite the success of PDR models in explaining the observations, we note that there are still limitations. From the model side, due to the clumpy and filamentary nature of the real molecular cloud, and the complex external/internal UV radiation field, the molecular abundance does not behave as expected from 1D models. From the observational side, there still have uncertainties in the calculation of molecular abundances, especially the sub-thermal excitation. We thus propose that the absorption measurements toward strong continuum sources (e.g., quasars, H\,{\sc ii} regions), in which the foreground gas usually has a simple density structure and uniform external UV field, could be useful to quantify the physical and chemical properties of the ISM.

\begin{acknowledgments}

We are grateful to the anonymous referee for the constructive comments and suggestions, particularly the suggestions of quantitative analysis on the CRIR, which have greatly improved this paper.
This work has been supported by the National Natural Science Foundation of China (grant No. 12041305), China Postdoctoral Science Foundation (grant No. 2021M691533), the Program for Innovative Talents, Entrepreneur in Jiangsu, the science research grants from the China Manned Space Project with NO.CMS-CSST-2021-A08.
TGB acknowledges support from Deutsche Forschungsgemeinschaft (DFG) grant No. 424563772.

This research has made use of the NASA/IPAC Infrared Science Archive, which is funded by the National Aeronautics and Space Administration and operated by the California Institute of Technology.
This work is based [in part] on observations made with the Spitzer Space Telescope, which was operated by the Jet Propulsion Laboratory, California Institute of Technology under a contract with NASA.
This publication makes use of data products from the Wide-field Infrared Survey Explorer, which is a joint project of the University of California, Los Angeles, and the Jet Propulsion Laboratory/California Institute of Technology, funded by the National Aeronautics and Space Administration.
Herschel is an ESA space observatory with science instruments provided by European-led Principal Investigator consortia and with important participation from NASA.
The James Clerk Maxwell Telescope is operated by the East Asian Observatory on behalf of The National Astronomical Observatory of Japan; Academia Sinica Institute of Astronomy and Astrophysics; the Korea Astronomy and Space Science Institute; Center for Astronomical Mega-Science (as well as the National Key R\&D Program of China with No. 2017YFA0402700). Additional funding support is provided by the Science and Technology Facilities Council of the United Kingdom and participating universities and organizations in the United Kingdom and Canada.
\end{acknowledgments}

%

\vspace{5mm}
\facilities{IRSA, Spitzer, WISE, Herschel, Arecibo, PMO:DLH, FCRAO, JCMT, FLWO:2MASS}


\software{Astropy \citep{Astropy2013,Astropy2018},  
          RADEX \citep{Van2007},
          emcee \citep{Foreman2013},
          {\sc 3d-pdr} \citep{Bisbas2012}
          }

\appendix

\section{Excitation temperature of CO}\label{sec:tex}

Here we use two strategies to derive the excitation temperature of CO:

\begin{itemize}

\item
$T_\mathrm{{ex}}$ derived from $^{12}$CO (1--0) ($T^1_\mathrm{{ex}}$).

The excitation temperature of CO can be derived from the intensity of optically thick $^{12}$CO lines ($T_\mathrm{12}$): $J(T_\mathrm{ex})=T_{12}+J(T_\mathrm{bg})$. To simplify the calculation, we adopt the maximum line intensity of the $^{12}$CO spectra as $T_\mathrm{12}$ at each position. 

We note that in the ECRs, the maximum intensity of $^{12}$CO can be as low as that of CMB, which may result in a negative value of $\tau_{13}$. In that case, we assume that the excitation temperature is 1\,K above the CMB.
\item
$T_\mathrm{{ex}}$ derived from $^{13}$CO (1--0) and (3--2) transitions ($T^2_\mathrm{{ex}}$).

With the assumption of LTE, the line ratio of the $^{13}$CO (3--2) and (1--0) transitions can be written as:
\begin{equation}
\begin{matrix}
\frac{\int T_{32} d\upsilon}{\int T_{10} d\upsilon} &= &3 \frac{x^2+x+1}{x^5} \frac{J(T_{ex})'-J(T_{bg})'}{J(T_{ex})-J(T_{bg})} , \\ &= &\frac{3}{x^5} \frac{15.866-0.048(x^3-1)}{5.288-0.89(x-1)}.
\end{matrix}
\label{eq:line_ratio}
\end{equation}
Where $x$ = $e^{h\nu_{10}/kT_\mathrm{ex}}$ = $e^{5.288/T_\mathrm{ex}}$.

Since the $^{13}$CO (3--2) observations only cover half of the region of $^{13}$CO (1--0), we thus adopt a mean $T_\mathrm{ex}$ derived above as the excitation temperature for the region that is uncovered by $^{13}$CO (3--2).
\end{itemize}

\begin{figure}
\includegraphics[width=1.0\linewidth]{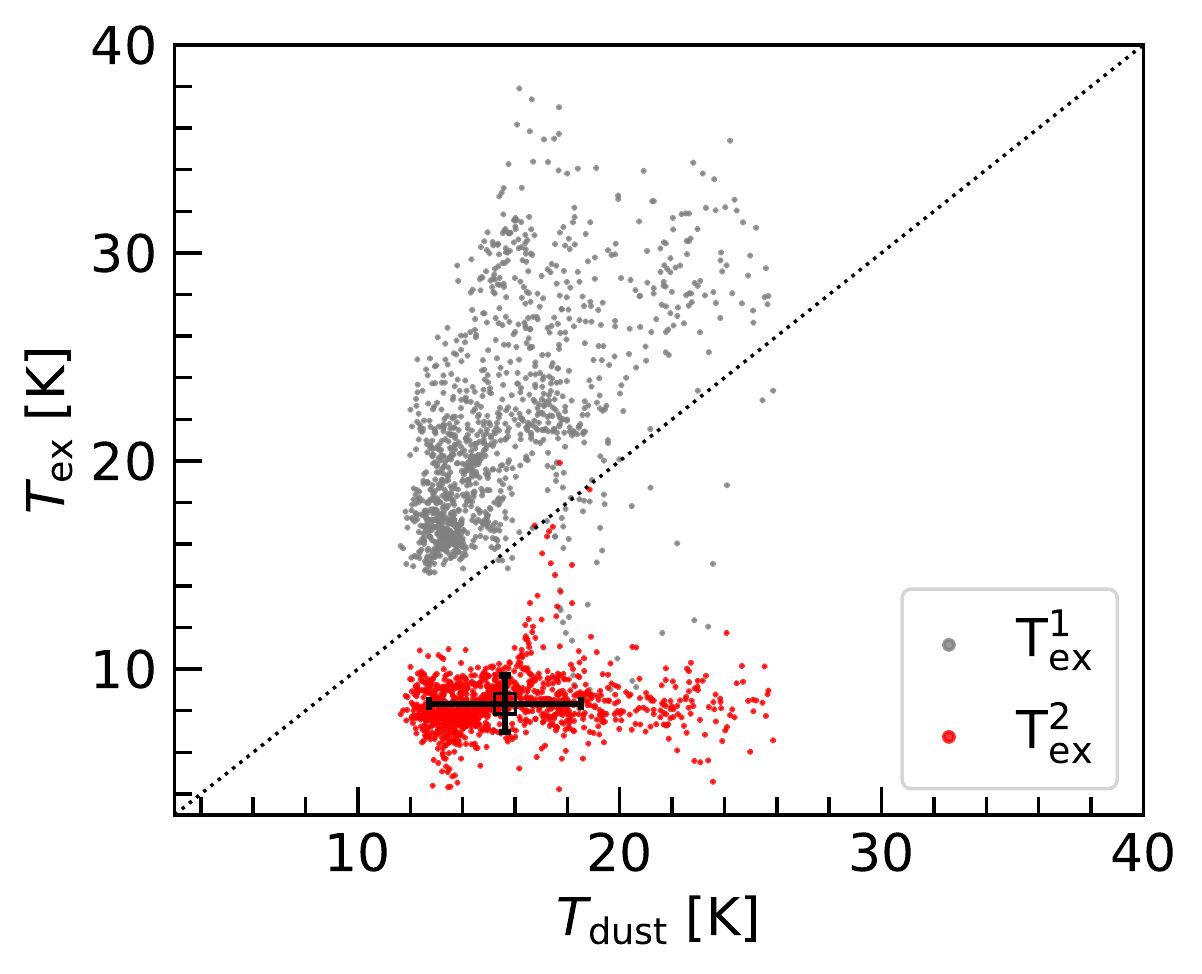}
\caption{Comparison of cold dust temperature ($T_\mathrm{d}$) and excitation temperature derived from $^{12}$CO (1--0) spectra ($T^1_\mathrm{{ex}}$, gray dots) as well as two transitions of $^{13}$CO (1--0) and (3--2) ($T^2_\mathrm{{ex}}$, red dots). The dash line represents $T_\mathrm{ex}$ = $T_\mathrm{dust}$. The black square with error bar denotes the mean value of $T_\mathrm{d}$ and $T^2_\mathrm{{ex}}$.}
\label{fig:tex}
\end{figure}

As shown in Figure \ref{fig:tex}, the excitation temperature of $^{12}$CO ($T^1_\mathrm{ex}$) is significantly higher than that of $^{13}$CO ($T^2_\mathrm{ex}$) and cold dust temperature ($T_\mathrm{d}$, see Section \ref{sec:uv}). The mean values of $T_\mathrm{d}$, $T^1_\mathrm{{ex}}$, and $T^2_\mathrm{{ex}}$ are 16$\pm$3\,K, 22$\pm$5\,K, and 8$\pm$1\,K, respectively. Since $^{12}$CO traces the outer layer of our molecular cloud while $^{13}$CO traces deeper zones, it means that the innermost regions may have a lower temperature than that of the outermost.
Numerical simulations find that a high far-UV (FUV) intensity increase the integrated intensity of CO by a factor of two or more \citep{Bisbas2021}. Observations in the Orion B molecular cloud also find overluminous CO gas due to the enhanced FUV field \citep{Pety2017}. Since IC~348 is located in the Per OB2 association, the FUV photons from nearby massive stars would excite the $^{12}$CO (1--0) transition at the outskirts of the molecular cloud.

Furthermore, the mean value of $T^2_\mathrm{{ex}}$ is much lower than the equivalent temperature of the upper energy level of $^{13}$CO (3--2) transition (31.73\,K), meaning that the transition of $^{13}$CO (3--2) is probably sub-thermally excited. More analysis of using non-LTE code to derive the column density is described in the Appendix \ref{sec:radex}.

\section{Column density of CO with RADEX}\label{sec:radex}

Given the specific input parameters (kinetic temperature, volume density of collision partners, column density of specific molecules, and linewidth), we can obtain a model spectra from the output of the non-LTE code RADEX \citep{Van2007}. In our simulation with RADEX, we consider H$_2$ as the collision partner, and the cold dust temperature ($T_1$ in Section \ref{sec:uv}) from the SED fitting as the kinetic temperature. 
The volume density of H$_2$ ($n_{\rm H}$) and the column density of $^{13}$CO ($N_\mathrm{^{13}CO}$) are free parameters, which are limited within specific parameter space (10$^1$ $<$ $n_{\rm H}$ $<$ 10$^6$\,cm$^{-3}$, 10$^{11}$ $<$ $N_\mathrm{^{13}CO}$ $<$ 10$^{19}$\,cm$^{-2}$). We then use the $emcee$ code \citep{Foreman2013} to maximize the likelihood function in the given parameter space, in which the likelihood function is defined as:
\begin{equation}
ln \ p = -\frac{1}{2} \sum_i \left [ \frac{\left ( F_{obs}^i - F_{model}^i \right )^2}{{\sigma^i_{obs}}^2} + ln \left ( 2\pi {\sigma^i_{obs}}^2 \right ) \right ],
\end{equation}
where the $\mathrm{F_{obs}^i}$ and $\mathrm{\sigma^i_{obs}}$ are the flux density and uncertainty of the observed $i$th transition, respectively, and $\mathrm{F_{model}^i}$ is the modelled flux density by RADEX. 


The best-fitted volume density given by RADEX is in the range of $\sim$ 10$^3$ to 2 $\times$ 10$^5$\,cm$^{-3}$, with a median value of $\sim$6 $\times$ 10$^3$\,cm$^3$. The inferred volume density is in good agreement with that of deriving from CS (2--1 and 3--2) transitions towards specific LOSs \citep{Baras2021}. The column density of $^{13}$CO is in the range of (0.38$\pm$0.03 $\sim$ 38.9$\pm$0.1) $\times$ 10$^{15}$\,cm$^{-2}$. Comparing with the results from $T^1_\mathrm{{ex}}$ in Section \ref{sec:co}, the column density of $^{12}$CO has reduced by a factor of 1.4$\sim$5 ($\sim$2 in average). Assuming that the $^{12}$C/$^{13}$C ratio is 65, the column density of $^{12}$CO ranges from (2.5$\pm$0.2) $\times$10$^{16}$\,cm$^{-2}$ to (2.53$\pm$0.01) $\times$10$^{18}$\,cm$^{-2}$.

\section{Extinction map overlaid with moment0 maps of $^{13}$CO and HCO$^+$}\label{sec:figures}

\begin{figure}
\includegraphics[width=1.0\linewidth]{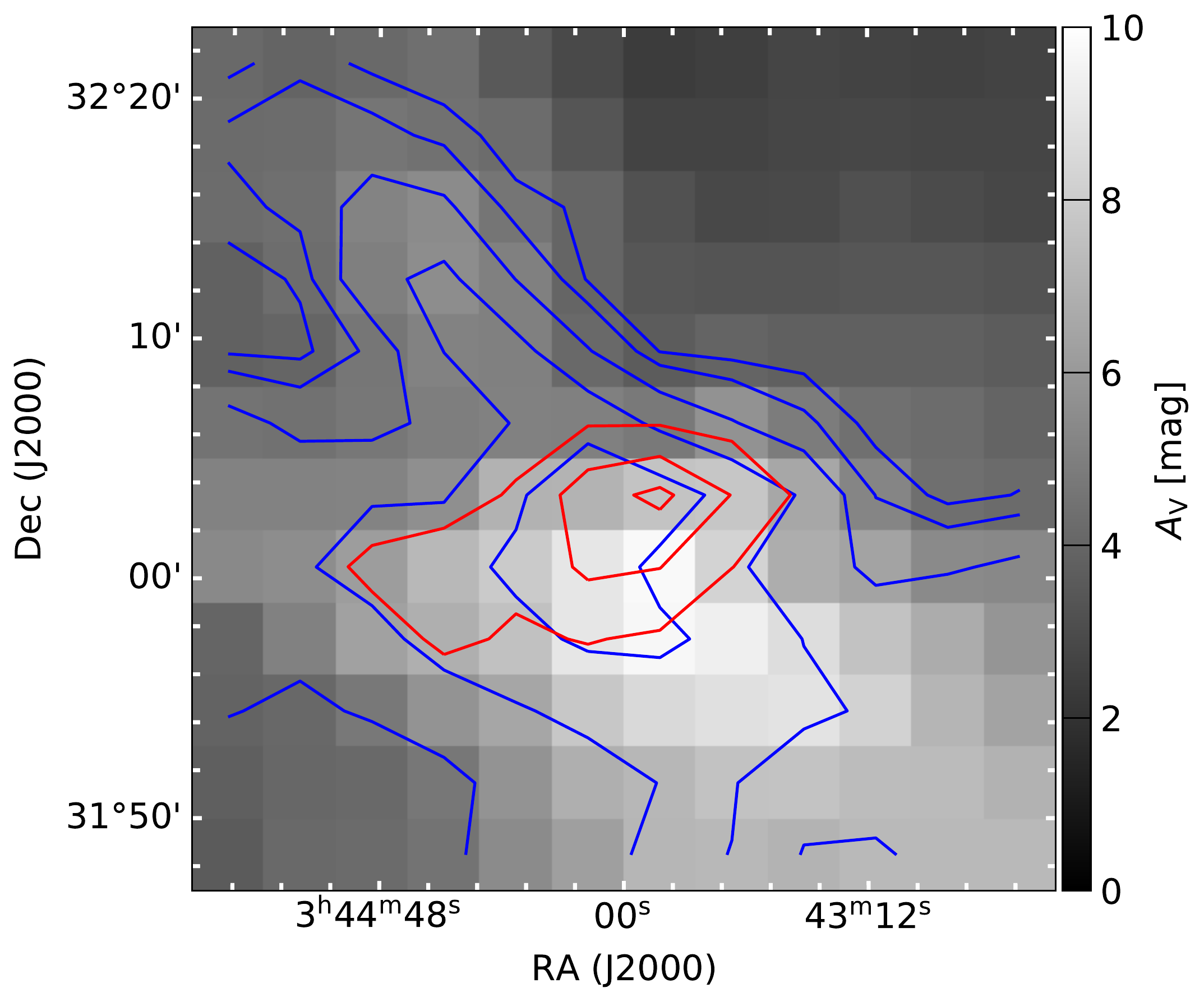}
\caption{The background image shows the extinction map from 2MASS. Blue and red contours denote the integrated intensity of $^{13}$CO and HCO$^+$, starting from 0.4\,K\,km\,s$^{-1}$ $\times$ (3, 5, 9, 15, 22) for $^{13}$CO and from  0.2\,K\,km\,s$^{-1}$ $\times$ (3, 5, 7) for HCO$^+$. All the images have re-binned to 3$'$ pixel size.}
\label{fig:per_image_rebin}
\end{figure}

We have re-binned the moment0 maps of $^{13}$CO (1--0) and HCO$^+$ (1--0) to match the resolution of the extinction map from 2MASS. Figure \ref{fig:per_image_rebin} shows the extinction map overlaid with the re-binned images.

\bibliography{reference}{}
\bibliographystyle{aasjournal}



\end{document}